\documentclass[10pt,aip,jcp,
amsmath,
amssymb,
preprint,
letterpaper,
				footinbib,%
				balancelastpage,
				raggedbottom, 
				superscriptaddress,
				citeautoscript
				]{revtex4-1}

\usepackage{graphicx}
\usepackage{dcolumn}
\usepackage{bm}

\usepackage[scaled=0.90]{helvet} 
\usepackage[protrusion=true,expansion=true,final]{microtype}
\usepackage{units,xspace}
\usepackage[usenames,dvipsnames]{xcolor}
\usepackage[bookmarks=false,colorlinks]{hyperref}
\hypersetup{
    linkcolor=magenta,        
    citecolor=OliveGreen,     
    filecolor=Plum,      	
    urlcolor=MidnightBlue,           
}
\usepackage{longtable}
\usepackage{tikz}

\usepackage{amsmath}
\usepackage{algorithm}
\usepackage[noend]{algpseudocode}

\begin{document}

\preprint{}
\title{Phase Classification of Multi-Principal Element Alloys via Interpretable Machine Learning}
\author{Kyungtae Lee}
	\affiliation{\footnotesize{Department of Materials Science and Engineering, University of Virginia, Charlottesville, VA 22904, USA}}%
\author{Mukil V. Ayyasamy}
	\affiliation{\footnotesize{Department of Materials Science and Engineering, University of Virginia, Charlottesville, VA 22904, USA}}%
\author{Paige Delsa}
	\affiliation{\footnotesize{Louisiana School for Math, Science, and the Arts, Natchitoches, Louisiana 71457, USA}}%
\author{Timothy Q. Hartnett}
	\affiliation{\footnotesize{Department of Materials Science and Engineering, University of Virginia, Charlottesville, VA 22904, USA}}%
\author{Prasanna V. Balachandran}
\email{pvb5e@virginia.edu}
	\affiliation{\footnotesize{Department of Materials Science and Engineering, University of Virginia, Charlottesville, VA 22904, USA}}%
	\affiliation{\footnotesize{Department of Mechanical and Aerospace Engineering, University of Virginia, Charlottesville, VA 22904, USA}}%
\date{\today}
\begin{abstract}
%
%
%
%
There is intense interest in uncovering design rules that govern the formation of various structural phases as a function of chemical composition in multi-principal element alloys (MPEAs).
In this paper, we develop a machine learning (ML) approach built on the foundations of ensemble learning, post hoc model interpretability of black-box models and clustering analysis to establish a quantitative relationship between the chemical composition and experimentally observed  phases of MPEAs.
%
The novelty of our work stems from performing instance-level (or local) variable attribution analysis of ML predictions based on the breakdown method, and then identifying similar instances based on $k$-means clustering analysis of the breakdown results.
We also complement the breakdown analysis with Ceteris Paribus profiles that showcase how the model response changes as a function of a single variable, when the values of all other variables are fixed.
Results from local model interpretability analysis uncover key insights into  variables that govern the formation of each phase.
%
Our developed approach is generic, model-agnostic, and valuable to \emph{explain} the insights learned by the black-box models. 
An interactive web application is developed to facilitate model sharing and accelerate the design of novel MPEAs with targeted properties.
%
%
%
\end{abstract}
%
\maketitle
\section{Introduction}
Multi-principal element alloys (MPEAs) are made by combining multiple elements, where every element contributes significant atom fraction to the alloy.\cite{senkov2015accelerated}
High entropy alloys (HEAs) represent a novel materials class within the broader family of MPEAs with outstanding mechanical, thermal, and electrochemical properties \cite{CANTOR2004213, HEA_2004, ZHANG20141, senkov_miracle_chaput_couzinie_2018, met6090199, Si_HEA_Radiation, CHEN201815, Miracle_Review_2014, doi:10.1002/adem.201700645, Miracle_2019_review, HEA_NRM_2019}. HEAs are unique amongst MPEAs because they  contain multiple (at least five) principal alloying elements of nearly equi-atomic concentration and yet have a global crystal structure with well-defined Bragg reflections indicative of long-range order. HEAs are typically solid solutions of face centered cubic (FCC), body centered cubic (BCC), or hexagonally closed packed (HCP) phases. Recently, the community has started to explore high entropic versions of intermetallic and ceramic compounds.\cite{HE_Ceramics, zhou2019single} To date, numerous elements in the periodic table have been explored to tune the properties of HEAs. However, not all compositions have resulted in the desired microstructure for application in extreme environments. In general, the physical and mechanical properties of HEAs vary depending on phase selection and their relative fractions in the microstructure\cite{WONG2018146, li_metastable_2016, CHEN2018129}.
In some applications, mixed phases are preferred\cite{tang2019designing}; whereas in others, a single-phase is desired.\cite{feuerbacher2018single}
Nonetheless, these observations have led many groups to develop effective and efficient phase prediction models for enabling discoveries of novel HEAs for targeted applications.

Traditional high-throughput approaches based on first-principles calculations are particularly not suitable to search for novel MPEAs due to the need for large supercells and complex crystal structure space involving multiple prototypes. Although computational thermodynamics based methods have played an important role,\cite{zhang2016calphad, feng2021high} their limitations are also documented in the published literature.\cite{Poon_HEA_SciRep} 
More recently, various groups have demonstrated the potential of data-driven machine learning (ML) methods to guide the design of MPEAs and HEAs towards promising regions in the search space\cite{ISLAM2018230, KIM2019124, Poon_HEA_SciRep, zhou_machine_2019, HUANG2019225, PhysRevMaterials.3.095005, QU2019299, KAUFMANN2020178, ZHANG2020108835, DAI2020109618, pei_npjCM2020, ZHANG2020528, RISAL2021110389, LEE2021109260, BENIWAL2021110647, YAN2021110723}. 

One of the most explored ML implementations on MPEAs research is the phase classification problem, where the objective is to train ML models for predicting whether a given chemical composition will form in: (1) single-phase FCC, BCC, or HCP solid solutions, (2) FCC+BCC dual phase with varying phase fractions, (3) single-phase intermetallics, (4) mixed phases (FCC+intermetalics, BCC+intermetallics, FCC+BCC+intermetallics, two different intermetallics etc.), or (5) amorphous phase.
%
%
ML models with fairly high accuracy (over 75\%) have been trained using small and large data set sizes and different choices of outputs.
Various elemental and thermodynamic properties have been considered as input features for the phase classification problem.\cite{ISLAM2018230, HUANG2019225, DAI2020109618, ZHANG2020528, KAUFMANN2020178, ZHANG2020108835, pei_npjCM2020}. 
%
A number of published studies also report descriptor importance based on cross-entropy, Gini index, or permutation methods to gain some insight into the descriptor contribution to the overall predictive power of the model. 
There are several drawbacks in the current
approaches. None of the published papers explain the predictions of the black-box models at the
granularity of each observation. There is a lack of principled approach to glean insights that shed light on the formation of each phase in the training set. It is not straight-forward to compare the predictive performance of every published ML
study using the data sets generated from different research groups, because the ML models
are not published along with the research paper.


\begin{figure}
   \begin{flushleft} (a) \end{flushleft}
   \vspace*{-3mm}
    \includegraphics[width=0.4\columnwidth]{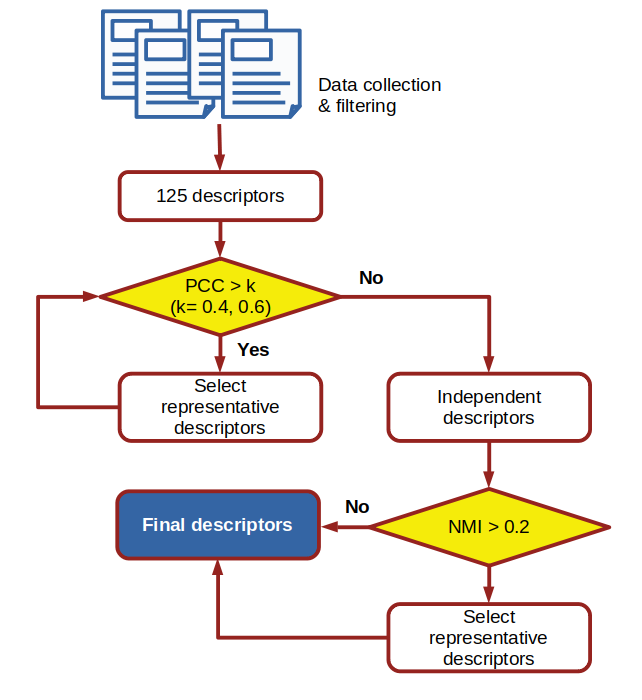}
    \begin{flushleft} (b) \end{flushleft}
   \vspace*{-3mm}
     \includegraphics[width=0.9\columnwidth]{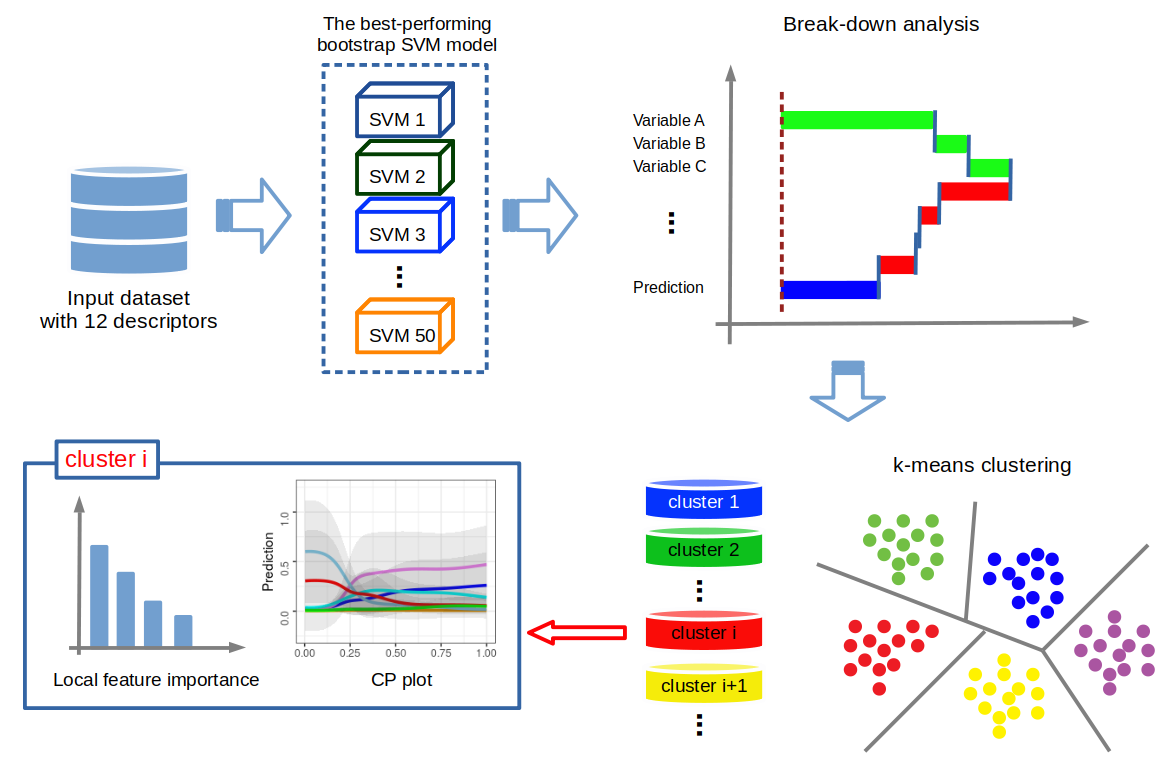}
    \caption{The flow chart for (a) feature selection using Pearson correlation coefficients (PCC) and normalized mutual information (NMI) and (b) machine learning and local model interpretability approach. In this work, we used an ensemble of support vector machines (eSVM) for multi-class classification learning, breakdown plots and Ceteris Paribus (CP) profiles for local model interpretability, and $k$-means clustering.}
 \label{fig:flowchart}
\end{figure}

In this work, we advance the application of ML methods in the MPEA phase classification problem in two significant ways. First, we apply two complementary instance-level (or local) post hoc model interpretability approaches, namely breakdown (BD) plots and Ceteris Paribus (CP) profiles, to glean insights into each observation. The BD method is based on the variable attribution principle, which decomposes the prediction of \emph{each individual observation} into particular variable contributions\cite{Staniak_2019}. In contrast, the more traditional global variable importance method provides a high-level or generic understanding of the inner workings of a black-box model and captures the relative importance of a given variable in impacting the overall model performance on the entire data set (that includes all phases). The CP profile method, on the other hand, evaluates the prediction response of a trained ML model to changes in a particular variable under the assumption that the values of all other variables do not change. We then develop a novel algorithm that combines the variable attribution data from the BD method with $k$-means clustering method to infer insights about similar instances. These results provide insight into explaining the relative variable contributions in the prediction of \emph{each phase or class label} as inferred by the ML models. In addition, the CP profile captures the average partial relationship between the predicted response and the input variables. 
In this paper, we demonstrate the power of local model interpretability methods as a key post hoc model analysis tool for materials informatics research. We apply them to explain the predictions from an ensemble of support vector machine (eSVM) models trained on a high-dimensional, multi-class MPEA phase classification problem data set. SVMs belong to a class of black-box models that lack transparency\cite{cortes_support-vector_1995, vapnik_estimation_2006}. More details about the eSVM approach is given in the Methods section.
%
Second, we build a novel interactive web application (\url{https://adaptivedesign.shinyapps.io/AIRHEAD/}) that allows the user to query our trained models directly and predict new MPEA or HEA compositions with the desired phase. This effort is aimed at allowing interested researchers to examine carefully the model predictions and facilitate the decision-making process. Moreover, this will also allow the MPEA community to objectively compare future models and document the progress. 

\section{Results}
\subsection{Data sets and Model training}
Our initial data set for ML was constructed by referring to several previous reports that meticulously compiled experimental data from the published literature. 
\cite{YANG2012233, hu_parametric_2017, SENKOV2016603, GUO201396, TODACARABALLO201676, Poon_HEA_SciRep, Miracle_ActMat2017, frozen_phonon, Gao_COSMS2017, Tan_JAC2018430, Ye_MT2016349, pei_npjCM2020, borg_SciData_2020}. The merged data set contained 3,719 compositions ranging from binary to multi-component alloys.
Each composition was also augmented with the phase information as reported in the literature.
The phases were then simplified into seven classes: BCC, FCC, BCC+FCC, HCP, Amorphous (AM), Intermetallics (IM), and Mixed-phases (MP). 
The IM label indicates that the microstructure contains at least one intermetallic phase.
The MP label indicates complex mixture of multiple phase combinations.
A final data set with 1,821 observations was obtained by removing all the duplicate data, missing values, and excluding the alloys showing inconsistent phase data depending on the source. 
Each of the 1,821 observation was represented by a total number of 125 variables.\cite{ward_general-purpose_2016, miedema_2005} We did not track the processing history, which can have an impact on the thermodynamics and kinetics of phase formation in the MPEAs.
%

The number of variables were then reduced based on linear Pearson correlation coefficient (PCC) \cite{PCC_2020} and non-linear normalized mutual information (NMI) analyses \cite{MI1177, NMI4749258}. The workflow is shown in Figure 1a. 
We considered two different PCC threshold values (0.4 and 0.6) to down-select least linearly correlated input variables. 
Our choice of using a PCC criterion of 0.6 was motivated by the work of Pei \emph{et al}.\cite{pei_npjCM2020}
In addition, we also imposed a more stringent PCC criterion of 0.4 for further simplification. 
%
The PCC analysis resulted in identifying 12 and 20 variable sets 
for the 0.4 and 0.6 criterion, respectively.
The list of down-selected variables is given in Table 1. We can broadly subdivide the down-selected variables in three categories: (1) those that are chemistry-agnostic (e.g., Mixing Entropy), (2) those that depend on element pairs (e.g., DeltaHf), (3) those that depend on chemistry (everything else in Table 1).
%

We then examined the presence of non-linear associations using NMI analysis, where an arbitrary criterion of NMI \textgreater\ 0.2 was adopted to flag the presence of any non-linear associations. The choice of 0.2 was informed by our calculated NMI value of 0.35 for a simulated sinusoidal curve.
While the 12 variable set showed no non-linear association, the 20 variable set contained three pairs of variables with NMI values greater than 0.2. We visualized the result using a scatterplot (Figure S1), which did not reveal any obvious non-linear trend that warranted further down-selection. Therefore, we ended up with two pre-processed data sets (one with a 12 variable set and the other with a 20 variable set) for ML model building. 
\begin{figure}
     \includegraphics[width=1.0\columnwidth]{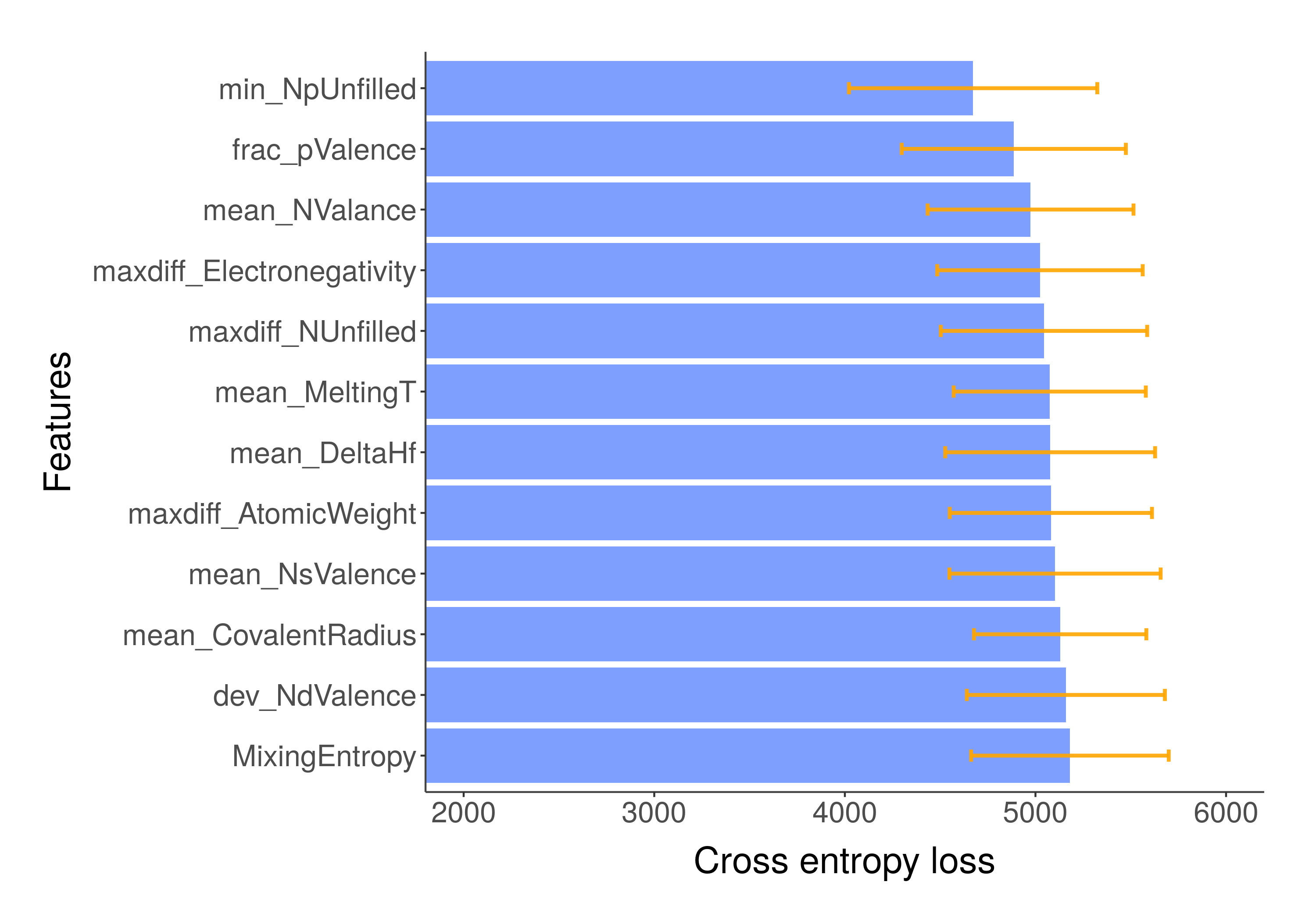}
    \caption{The global variable importance for the 12 variable set eSVM model. Cross entropy loss was used as an indicator of variable importance. Error bars represent the standard deviation from 50 SVM models in the ensemble.}
 \label{fig:space}
\end{figure}

\begin{table*}
\caption{\label{tab:representation}List of the descriptors identified from 125 descriptors by PCC \textgreater\ 0.4 or 0.6 along with NMI \textgreater\ 0.2.}
\footnotesize\rm
\begin{center}
\begin{tabular}{p{4cm} p{3cm} p{9cm}}
\hline
Notation & PCC & Description\\
\hline
maxdiff\_NUnfilled & \textgreater\ 0.4 only & Difference between minimum and maximum numbers of unfilled valence orbitals \\
min\_NpUnfilled & & Minimum number of unfilled $p$ valence orbitals \\
\hline
dev\_NsValence & \textgreater\ 0.6 only & Standard deviation of the number of filled $s$ valence electrons \\
dev\_CovalentRadius & & Standard deviation of covalent radius \\
dev\_NdUnfilled & & Standard deviation of the number of unfilled $d$ valence orbitals \\
dev\_NUnfilled & & Standard deviation of the number of unfilled valence orbitals \\
mean\_NUnfilled & & Average number of unfilled valence orbitals \\
dev\_NpUnfilled & & Standard deviation of the number of unfilled $p$ valence orbitals\\
maxdiff\_MeltingT & & Difference between minimum and maximum melting temperatures\\
variance\_DeltaHf & & Standard deviation of mixing enthalpy \\
min\_NpValence & & Minimum number of filled $p$ valence electrons \\
min\_NdUnfilled & & Minimum number of unfilled $d$ valence orbitals \\
\hline
maxdiff\_AtomicWeight & Both & Difference between minimum and maximum atomic weights\\
mean\_NValance & & Average number of filled valence electrons \\
mean\_MeltingT & & Average melting temperature \\
mean\_NsValence & & Average number of filled $s$ valence electrons \\
dev\_NdValence & & Standard deviation of the number of filled $d$ valence electrons\\
frac\_pValence & & Fraction of filled $p$ valence electrons \\
MixingEntropy & & Mixing entropy \\
mean\_DeltaHf & & Average mixing enthalpy \\
maxdiff\_Electronegativity & & Difference between minimum and maximum electronegativity values \\
mean\_CovalentRadius & & Average covalent radius of constituent elements \\

\hline
\end{tabular}
\end{center}
\end{table*}

The pre-processed data set was randomly split into two subsets with 75\%  and 25\% data for training and testing, respectively. 
We used the eSVM algorithm for training the ML models. The optimal hyperparameters were determined using a grid search. The out-of-bag error rate was used to evaluate the performance. We systematically varied the number of bootstrap samples and found the 50 and 100 bootstrap eSVM models to show the best predictive performance on the test data for the 12 and 20 variable sets, respectively.
%
Tables S1 and S2 compare the relative performance of eSVM models on the test set in terms of accuracy, precision, recall, and F1-score.
Both 12 and 20 feature sets of eSVM showed similar performance.
%
%
%
Finally, we chose the simpler 12 feature set eSVM models for further analysis.
The next step is the post hoc analysis of the trained eSVM models. We start with the global variable importance analysis, which is also the most common method within the ML MPEA community.
%

\subsection{Global Variable Importance}
The objective of global variable importance analysis is to evaluate the relative importance of each variable in impacting the overall predictive performance of the trained ML models. In this work, we used the well known permutation-method and cross-entropy loss function to assess the global variable importance.\cite{Dalexpackage} In Figure 2,  we show the averaged global variable importance analysis from the 12 feature set eSVM model. All features appear to contribute to the prediction performance of the eSVM model. The error bar is the standard deviation from the 50 bootstrap samples. Mixing entropy, number of filled $d$ or $s$ valence electrons, covalent radius, and atomic weight are identified as more important to affect the prediction performance. This result agrees well with the various ML papers in the literature\cite{ISLAM2018230, HUANG2019225, KAUFMANN2020178, ZHANG2020108835, DAI2020109618, ZHANG2020528}.
%
While helpful, global variable importance approach does not shed light on the following question: \emph{what variables contribute to the prediction of each phase (or class label) and how are these variables related to the predicted phase?} This requires an implementation of local variable importance methods, which we discuss next.
%

\subsection{Local Variable Importance}
We focused on two complementary local model interpretability methods: (1) Breakdown plots and (2) Ceteris Paribus profiles.

\subsubsection{Breakdown analysis}
\begin{figure}
     \includegraphics[width=1.0\columnwidth]{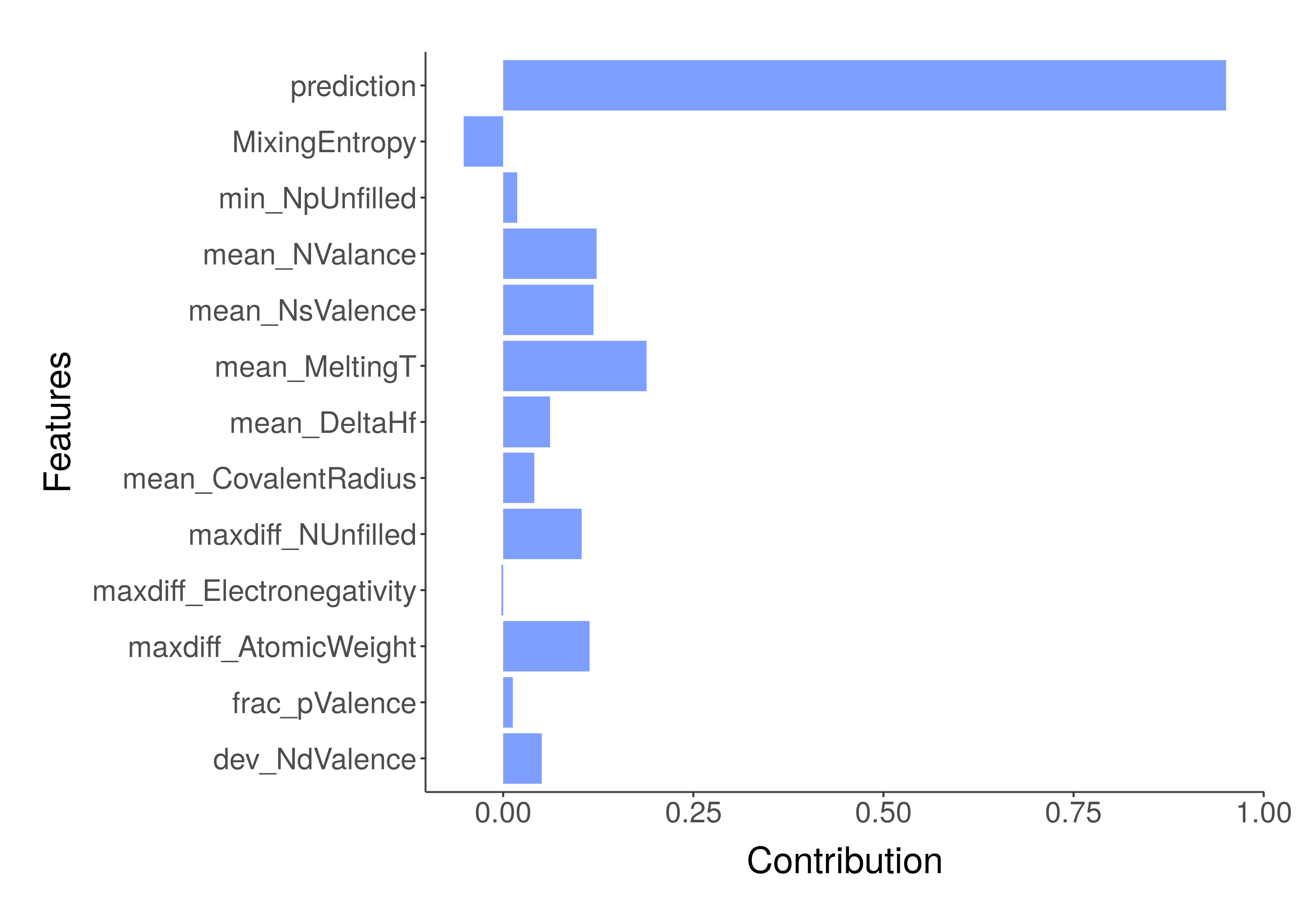}
    \caption{The BD plot for NbTaTiV composition, which is predicted to form in BCC phase by the eSVM model. Each bar represents the averaged contribution for that variable towards the overall prediction.}
 \label{fig:space}
\end{figure}

In the breakdown (BD) approach, we decompose the model prediction for a single observation into contributions that can be attributed to different input variables.\cite{Dalexpackage, RJ-2018-072} The BD analysis can start from either a null set of indexes or a full set of relaxed features, which are referred to as step-up and step-down approaches, respectively. In the case of step-down approach (as considered in our work), each contribution of input variable is calculated by sequentially removing a single variable from a set followed by variable relaxation in a way that the distance to the prediction is minimized. 
For example, in Figure 3, a BD plot is shown for the NbTaTiV composition. The eSVM model predicted the composition to form in BCC with 100\% probability score. Thus, we will obtain only one BD plot for this composition representing the BCC phase prediction. The BD plots resemble a bar graph. Each variable can either contribute positively (positive weight) or negatively (negative weight) to the overall prediction. In this specific example, the mean\_MeltingT, mean\_NValence, and mean\_NsValence variables carry the largest weight and are recognized as important for predicting the composition as forming in the BCC phase. In a similar manner, we calculated the BD plots for all compositions in the training data. Readers can access the BD plots through our Web App.

\subsubsection{Ceteris Paribus profile}

\begin{figure}
     \includegraphics[width=1.0\columnwidth]{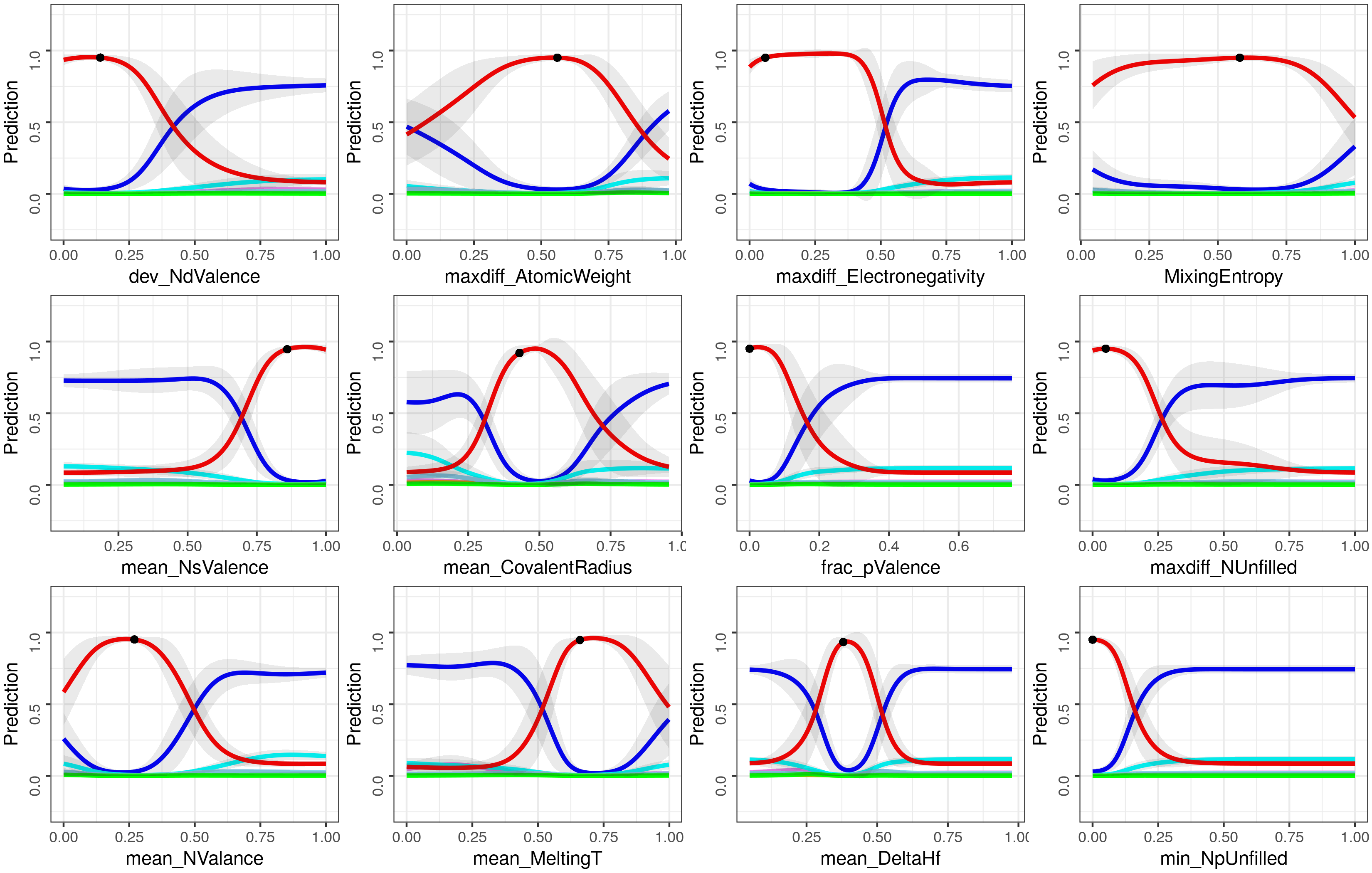}
    \caption{The CP profile for NbTaTiV composition with respect to the 12 input variables. The black dots indicate the true feature values. Line colors denote phase information: blue, MP; violet, AM; cyan, FCC; orange, BCC+FCC; lightblue, HCP; red, BCC; green, IM.}
 \label{fig:space}
\end{figure}

%

The Ceteris Paribus (CP) profiles convey complementary insights about the relationship between a variable and the response by showing how the prediction would be affected if we changed a value of one variable while keeping all other variables unchanged.\cite{Dalexpackage} The method is based on the Ceteris Paribus principle; ``Ceteris Paribus'' is a Latin phrase meaning ``other things held constant'' or ``all else unchanged.'' CP profiles are an intuitive method to gain insights in to how the black-box model works by investigating the influence of input variables separately, changing one at a time.\cite{Dalexpackage} In essence, a CP profile shows the dependence of the conditional expectation of the dependent (or output) variable on the values of a particular input variable. In Figure 4, we show a representative CP profile plot for the same NbTaTiV composition that was discussed in the previous BD section. Unlike the BD plot, we also observe the functional dependence of each variable on the model performance. In Figure 4, x-axes are the input variables and the y-axes are the prediction probabilities from the eSVM models. There are seven curves in each panel and each curve represents a particular phase. For example, the red curve traces the prediction for the BCC phase. The CP profile plot highlights the presence of non-linear relationship between each of the feature and the response. CP profiles for other compositions can be accessed through our Web App.

\begin{figure*}[htp]
     \includegraphics[width=0.95\columnwidth]{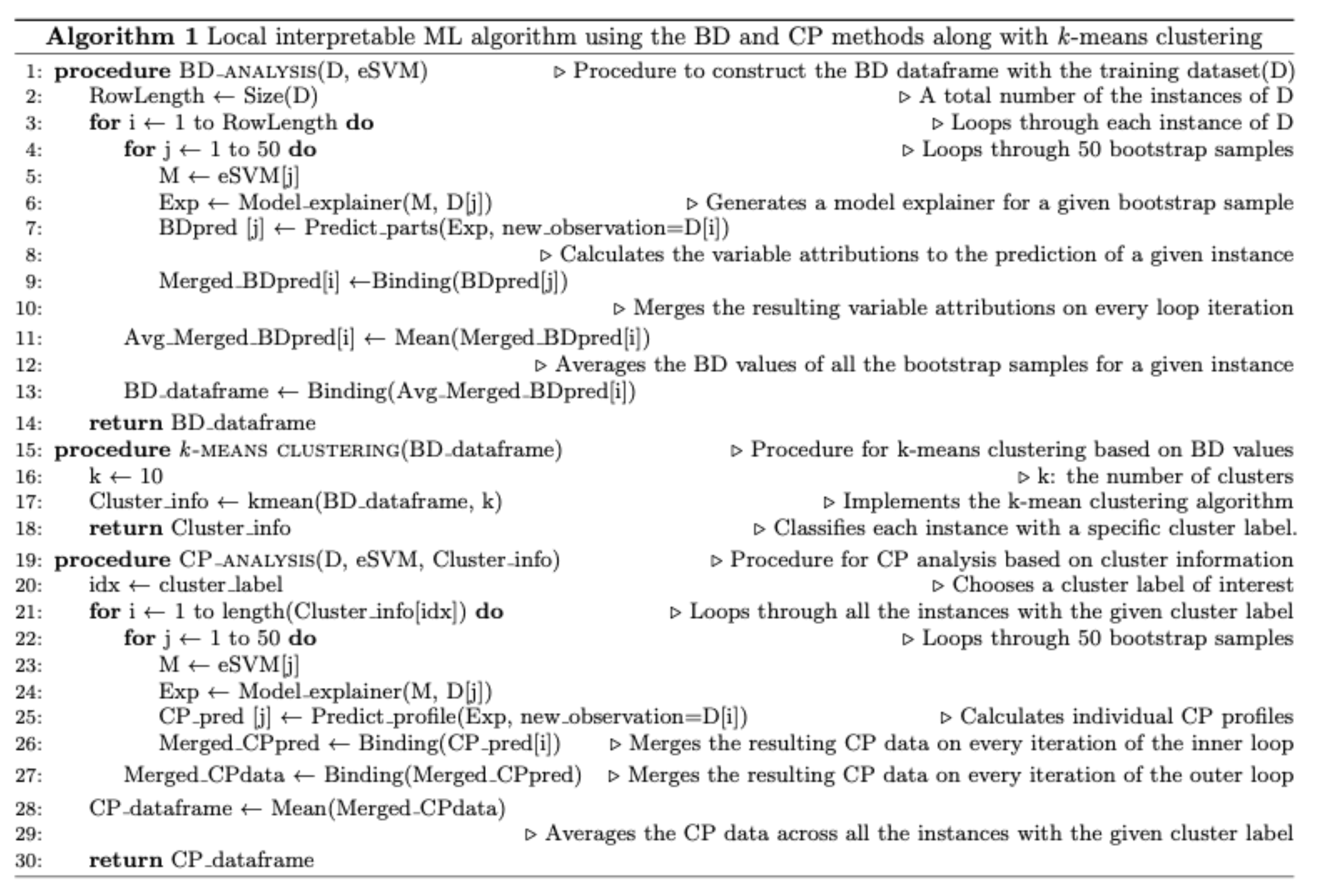}
 \label{fig:algorithm}
\end{figure*}

\subsection{Extracting Variable Importance for each Phase}
While the global variable importance analysis functions at the entire data set level, the breakdown and Ceteris Paribus analyses function at the granularity of each instance or composition. These two methods represent the two extremes in the spectrum of post hoc model interpretability analysis. In addition, there is a need for model interpretability at the \emph{intermediate level} that will yield insights specific to each phase in our data set (based on the collective similarity or clustering of similar observations).
To address this question, we combined the BD plots with the $k$-means clustering analysis and CP profile data. The pseudocode is summarized in Algorithm 1, which describes the implementation sequence of the BD method, \emph{k}-means clustering, and CP analysis. 
%

The algorithm starts with the BD analysis for each composition.
%
For a given composition, the BD values are calculated from each trained SVM model in the ensemble and averaged across all 50 ensembles. The results are stored as a data frame. We then perform clustering analysis using the \emph{k}-means algorithm, assigning a cluster label to each data point. We also construct CP profiles for each composition in the data set and group them according to the cluster labels. We then calculate the average CP profile for each cluster. The final outcome is two plots for each cluster: (1) averaged BD plots and (2) averaged CP profiles. Visualization of the two plots will yield phase-specific interpretation of the eSVM model.
For \emph{k}-means clustering, we found the optimal number of clusters by plotting the total within sum of square as a function of the number of clusters (Figure S2a). The elbow point corresponded to the choice of 10 clusters (as visualized in Figure S2b using principal component analysis).
The 10 clusters were then analyzed using histograms as shown in Figure 5, where we plot the frequency of occurrence of the number of components in the alloy composition for each cluster. 
Figure 5 shows that clusters 1, 5, 7 and 10 capture patterns that are representative of the binary systems. 
Given our interest in the design of HEAs, which normally consists of more than four components, we do not discuss the results from clusters 1, 5, 7, and 10. All other clusters can provide important clues for uncovering phase-specific variable importance analysis that pertain to the MPEAs and HEAs. Instead of explaining each cluster in detail (which is beyond the scope of this paper), we only focused on specific clusters where the ML predictions agreed closely with the experimental labels in the data set.


\begin{figure}
     \includegraphics[width=1.0\columnwidth]{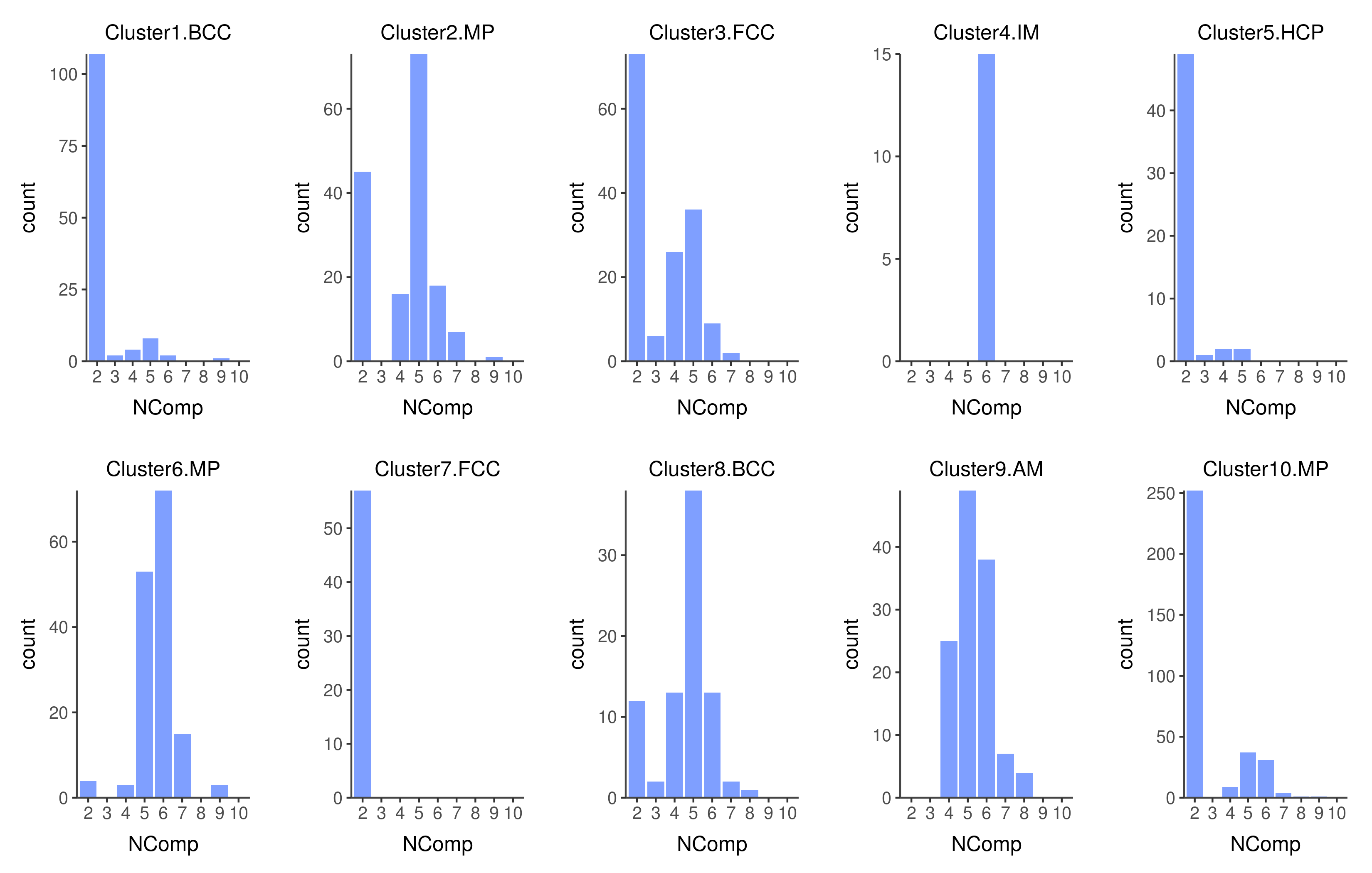}
    \caption{The distributions of the number of components (denoted as NComp) for the 10 clusters from $k$-means clustering analysis. Each cluster is also identified by phase selections via the BD-based prediction as shown in the titles of each plot.}
 \label{fig:space}
\end{figure}


\begin{figure}
     \begin{flushleft} (a)  \end{flushleft}
     \vspace*{-3mm}
     \includegraphics[width=0.7\columnwidth]{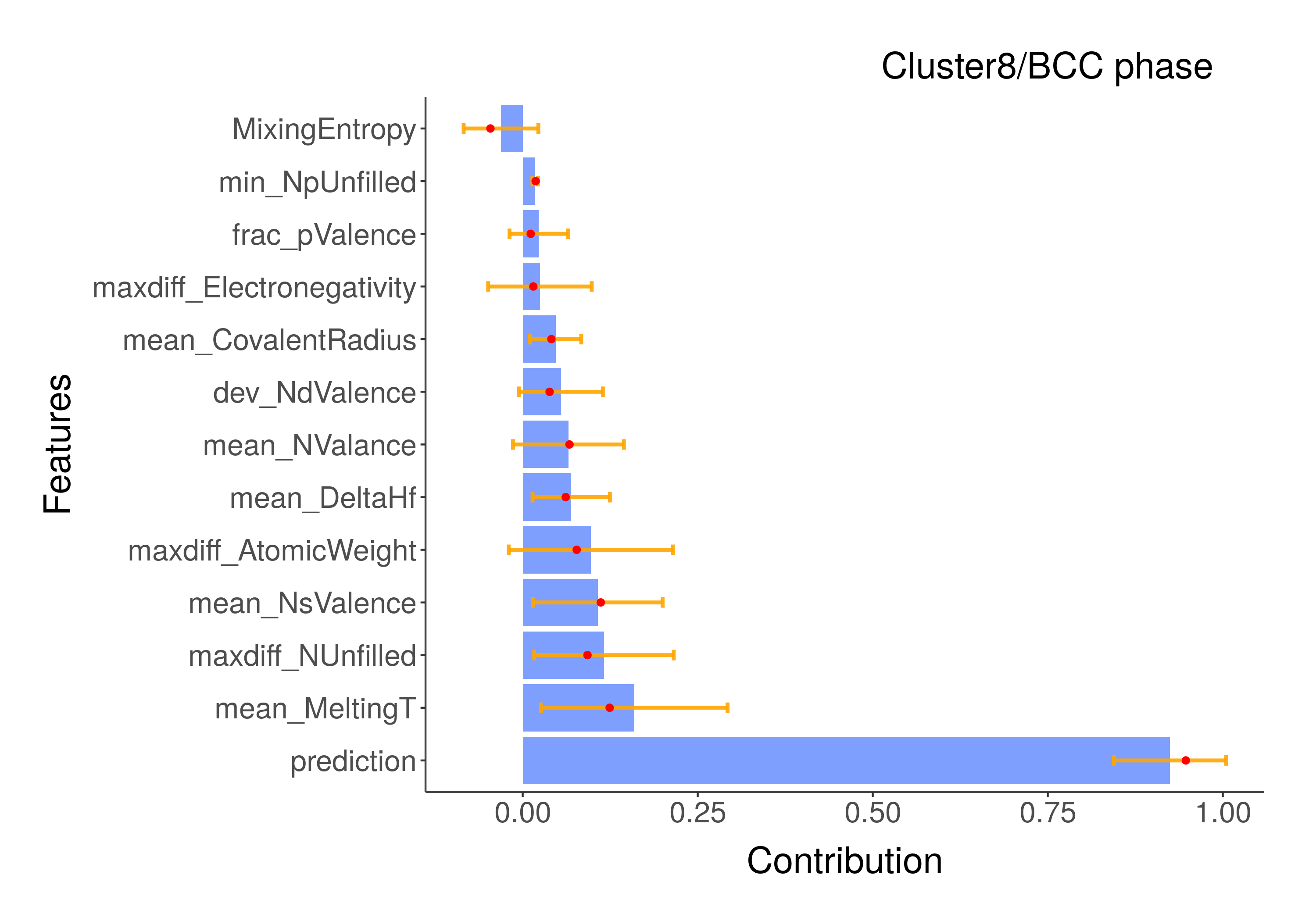}
     \begin{flushleft} (b)  \end{flushleft}
     \vspace*{-3mm}
     \includegraphics[width=0.7\columnwidth]{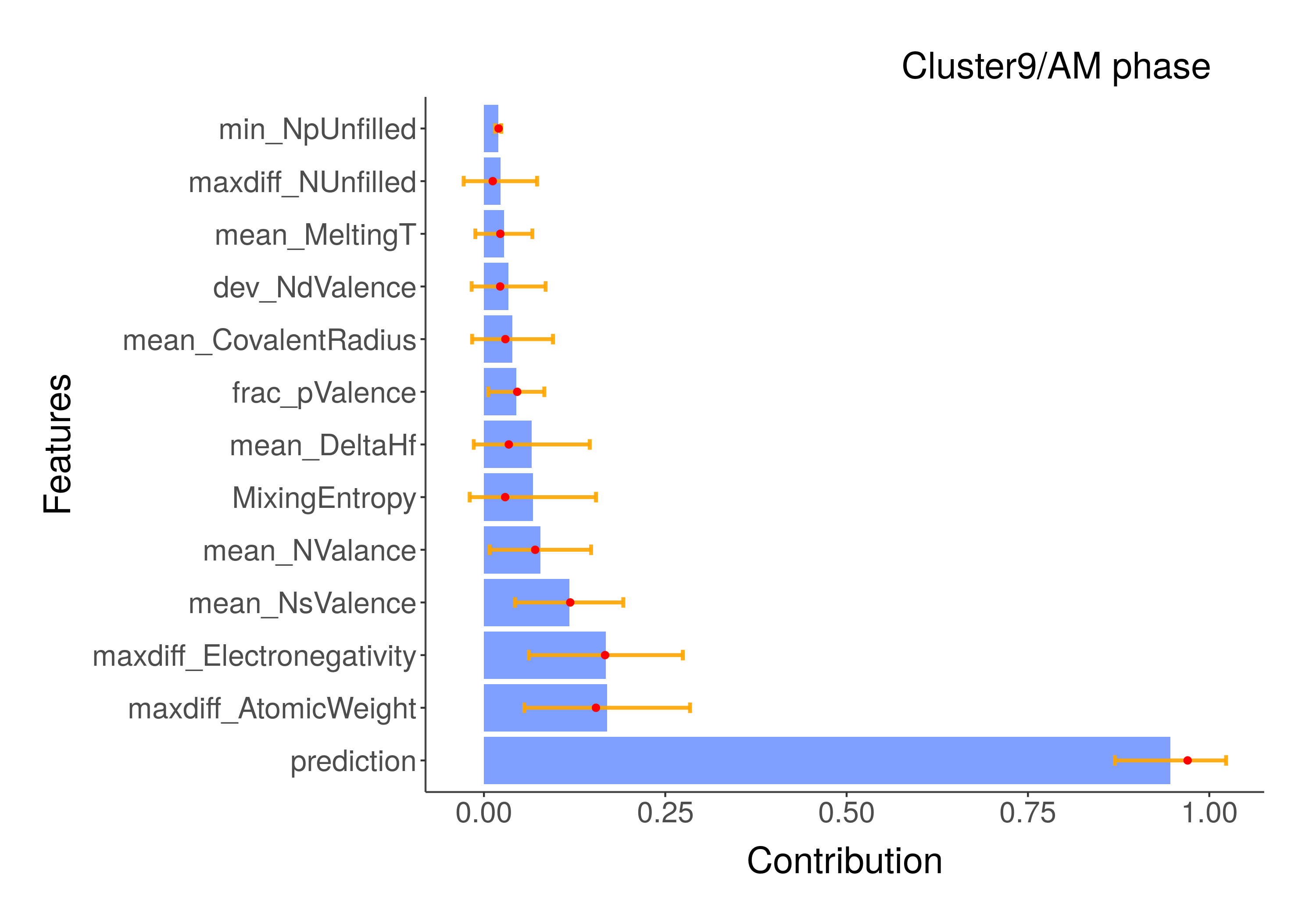}
    \caption{The averaged and sorted contribution from each variable for (a) cluster 8 (BCC phase) and (b) cluster 9 (AM phase). Each bar represents the relaxed predictions with and without a particular single explanatory variable in the corresponding row. The last row contains the sum of the overall mean prediction values. Red dots and yellow lines stand for median values and error bars, respectively.}
 \label{fig:space}
\end{figure}

In Table S3 (in the Supplemental Document), we compared the ML prediction accuracy for each of the 10 clusters. 
Figure 5 indicates that clusters 8 and 9 are representative of the MPEAs. Although cluster 4 is also representative of MPEAs (six-component alloys), it contained fewer data points than clusters 8 and 9. Therefore, we focused on clusters 8 and 9 for model interpretation.
The prediction accuracy data from eSVM reveals that clusters 8 and 9 are representative of the BCC and AM phases, respectively.
The averaged variable attribution analyses from the BD method for clusters 8 and 9 are shown in Figures 6a and b, respectively. The mean\_NsValence and maxdiff\_AtomicWeight variables are identified as important variables for both BCC and AM phases. Since the maxdiff\_AtomicWeight variable can be related to the atomic size mismatch, this result is in good agreement with the previous studies.\cite{20001372Takeuchi, 2005Takeuchi}
%
Figure 6a indicates that mean\_MeltingT, maxdiff\_NUnfilled, and mean\_DeltaHf are key variables for the formation of BCC phase. From Figure 6b, it can be inferred that  maxdiff\_Electronegativity, mean\_NValence, and MixingEntropy are important for forming the AM phase. The relationship between mean\_DeltaHf and BCC phase also agrees well with the previous published results.\cite{200700240Zhang}

The averaged BD plots from other clusters are also displayed in Figure S3, and the interpretations are summarized in Table S4. 
The analysis reveal similarities between BCC and IM phases, and between FCC and AM phases. The MP phase does not appear to have distinct characteristics. This may be due to the fact that the alloys of MP phase have a wider range of data distribution arising from relatively more abundant data and many different types of mixed phases compared to those with other phases that are more unique.

We next visualize the averaged CP profiles for clusters 8 and 9, which provide a more detailed account of the relationship between the input variables and the phases. The CP profiles for BCC and AM phases are shown in Figures 7a and b, respectively. Not all input variables have unique functional relationships. For example, in Figure 7a (representative of BCC phase), similar functional relationships are observed between: (1) frac\_pValence, maxdiff\_NUnfilled and min\_NpUnfilled, (2) mean\_CovalentRadius and mean\_DeltaHf, (3) dev\_NdValence, maxdiff\_Electronegativity, and mean\_NValance, and (4) mean\_NsValence and mean\_MeltingT. The maxdiff\_AtomicWeight and MixingEntropy are the only two variables that do not share a similar relationship with any other variable. 

We also made an attempt to connect the averaged BD plots (Figure 6a) with the averaged CP profiles (Figure 7a) for the BCC phase. We found that High mean\_MeltingT, high mean\_NsValence, and mean\_DeltaHf values between 0.3 and 0.5 favor BCC phase formation. From the standpoint of 
maxdiff\_AtomicWeight and maxdiff\_NUnfilled variables, MPEAs tend to form in BCC phase when the constituent elements have moderately different atomic weights and similar number of the unfilled valence orbitals. 
In the case of AM phase (Figure 7b), while high mean\_NsValence values are preferred, low mean\_NValence values favor AM phase formation. Low MixingEntropy should be avoided, because it appears to favor the formation of mixed phase (blue curve in Figure 7b). There is a window of values for maxdiff\_AtomicWeight and maxdiff\_Electronegativity that favor AM phase formation. In Figure 7b, extreme values of maxdiff\_AtomicWeight and maxdiff\_Electronegativity appear to favor mixed phase.

\begin{figure}
     \begin{flushleft} (a)  \end{flushleft}
     \includegraphics[width=0.8\columnwidth]{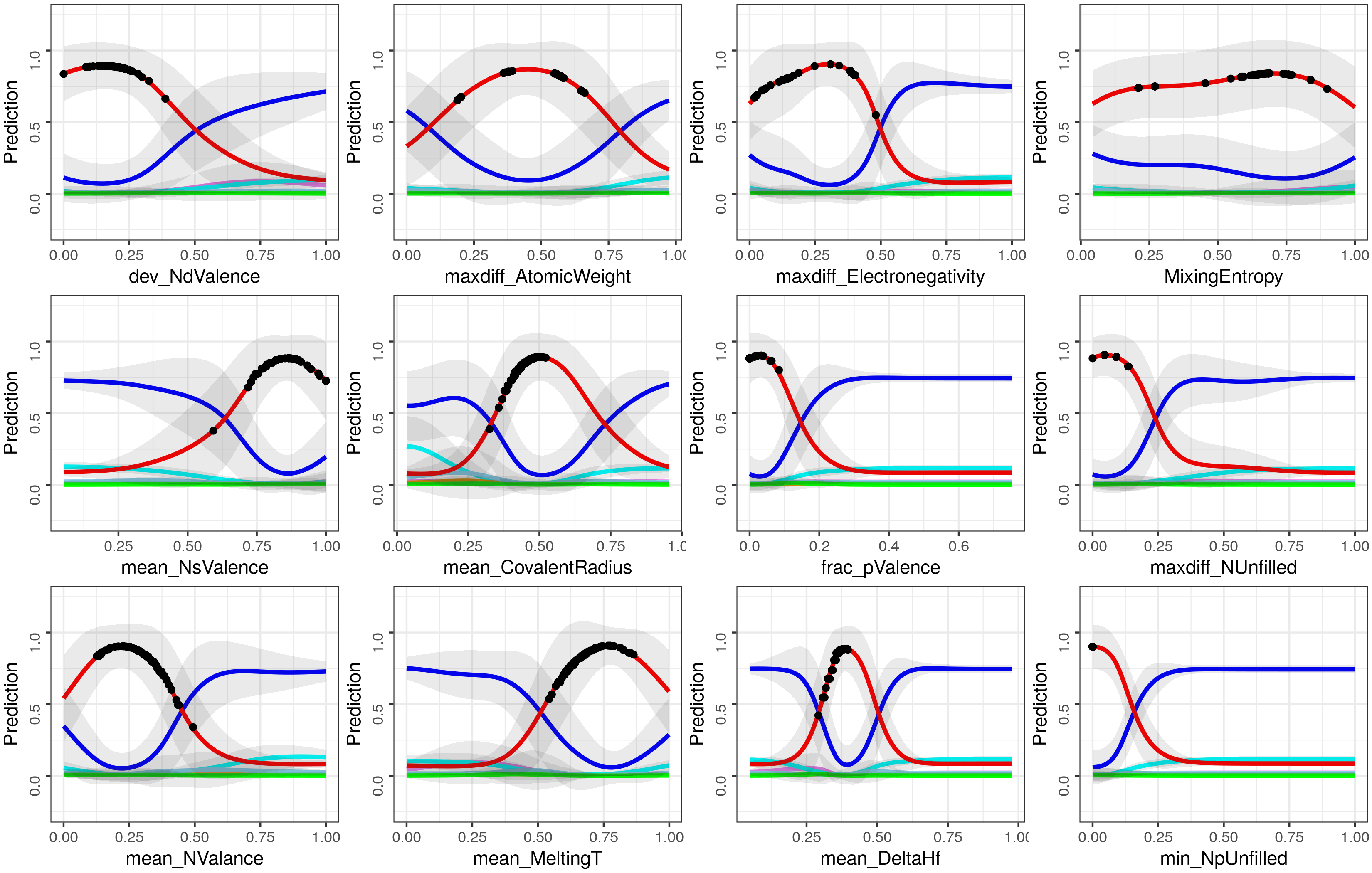}
     
     \begin{flushleft} (b)  \end{flushleft}
     \includegraphics[width=0.8\columnwidth]{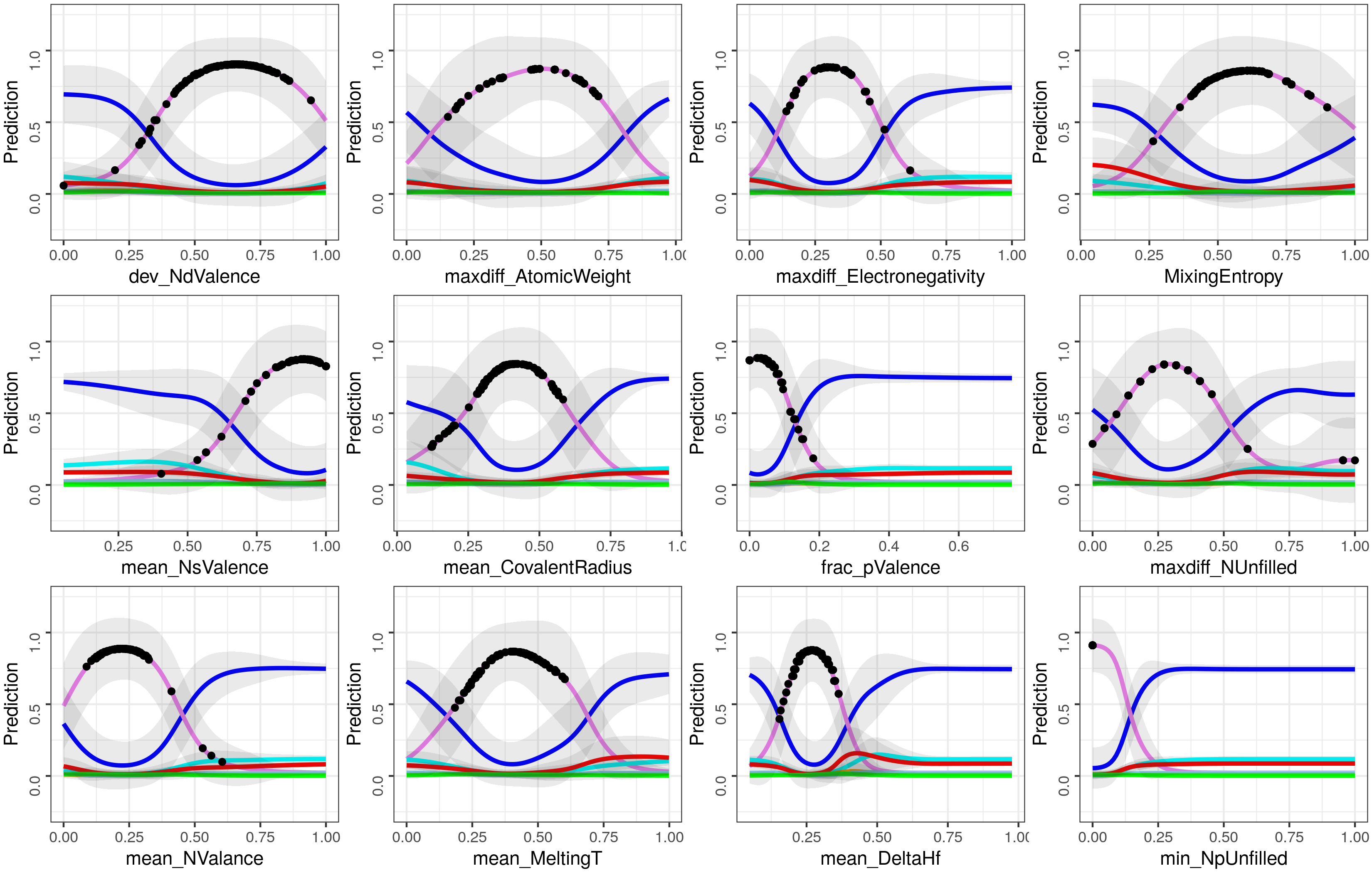}
    \caption{The averaged CP profiles for (a) cluster 8 (BCC phase) and (b) cluster 9 (AM phase) with respect to the 12 input variables. The black dots indicate the true feature values for all the data points within that cluster. Line colors denote phase information: blue, MP; violet, AM; cyan, FCC; orange, BCC+FCC; lightblue, HCP; red, BCC; green, IM.}
 \label{fig:space}
\end{figure}

So far, we have been comparing the averaged CP profiles within a cluster. We also observe some interesting patterns between the two clusters. For example, maxdiff\_AtomicWeight, mean\_CovalentRadius, mean\_NValance, mean\_NsValence,  frac\_pValence, mean\_DeltaHf and min\_NpUnfilled have similar functional forms. In contrast dev\_NdValence, maxdiff\_Electronegativity, MixingEntropy, maxdiff\_NUnfilled and mean\_MeltingT show distinct functional dependencies. The implications of these results are not entirely clear, but showcases the potential of local model interpretability methods for in-depth examination of the black-box models.

In Figure 8, we show the distribution of constituent elements in clusters 8 and 9.
The elements on the left side of the $d$-block in the periodic table, along with Al, are found in the BCC cluster (cluster 8). In contrast, the compositions representing the AM phase (cluster 9) show a scattered distribution of elements from the $d$-block. The existence of Be atom in the AM cluster likely implies the connection between the AM phase and a large difference in atomic weight. From the pie charts, we can see that both Ti and Zr are the major elements in both BCC and AM clusters. When it comes to unique elemental constituents, the elements of Nb, Ta, Mo, and V are commonly found in the BCC phase, whereas Cu, Ni, and Al are in the AM phase. Other clusters are also analyzed in the same manner and the results are shown in Figure S4. For FCC, the constituent elements are distributed in the first and second rows of the $d$-block from the periodic table. 
The MP phase is similarly related to the first row of the $d$-block, but several of the $p$-block elements also participate in the formation of MP phase. 

\begin{figure*}[htp]
     \includegraphics[width=0.95\columnwidth]{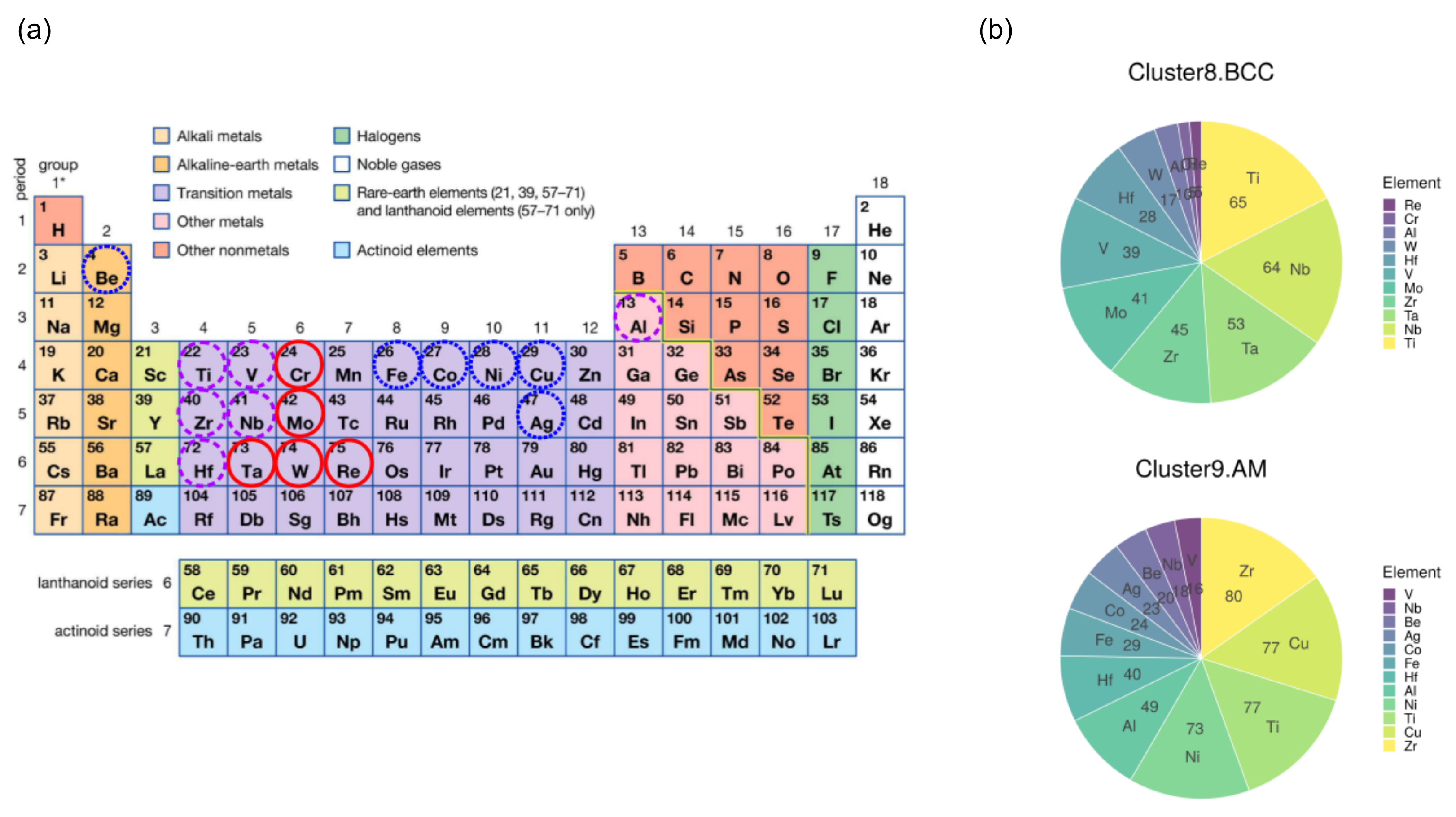}
    \caption{The constituent elements present in clusters 8 (BCC phase) and 9 (AM phase) are (a) depicted in the periodic table and (b) analyzed by pie charts, where each number shows their frequency of occurrence. The purple (dashed), red (solid), and blue (dotted) circles indicate the elements appearing in both BCC and AM phases, only BCC phase, and only AM phase, respectively.} 
 \label{fig:space}
\end{figure*}


\section{Discussion}
There is an increasing interest in the application of model interpretability tools to problems in materials science.\cite{ling2017building, xie2018crystal, lopez2017design, Rampi_SHAP, XIONG2021133}
The expectation that the ML model should also explain the underlying patterns of materials phenomena in addition to the predictions has been steadily increasing. 
There are also papers from other disciplines, such as bioinformatics, that share similar goals \cite{clough2019global}.
We have developed a novel post hoc ML model interpretability framework for the MPEA phase classification problem.
The algorithms provide an in-depth analysis of the complex black-box models and extracts interpretable patterns from an ensemble of trained models.
In the materials informatics literature, the results from global variable importance are widely used to interpret which variables are strongly related to the ML performance. 
We argue that phase-specific (or class label specific) variable importance analysis based on local model interpretability offers a new way to gain much deeper insights into the global variable importance results.
%
To illustrate this point, we also compared the global and local variable importance plots to glean additional insights (main results are distilled in Table S5).
Note that the top three variables from the global variable importance analysis, namely MixingEntropy, dev\_NdValence, and mean\_CovalentRadius, are not associated with either the single-phase BCC or FCC compositions that have attracted interest for tailoring the mechanical properties of the HEAs.\cite{GEORGE2020435}
The fact that these variables are connected to the MP phase indicates that the presence of a large fraction of the MP phase in the dataset significantly affects (or biases) the global variable importance analysis. 
One can also find that the important variables for BCC and FCC from the BD plots are not ranked highly by the global variable importance. 
Therefore, pursuing MPEA design based solely from global variable importance analysis could potentially mislead the researchers especially from the context of a multi-class classification learning setting.
Augmenting global variable importance analysis with local feature importance has many desirable characteristics for rationally tailoring new HEAs with desired properties. 
%
%

\section{METHODS}
\noindent \textbf{Data preprocessing.} The dataset collected from the literature consists of 1,821 compositions after deleting the duplicate data and missing values. Descriptors are generated by the Magpie program \cite{ward_general-purpose_2016} which is a package to compute the concentration-weighted values of materials using the elemental or pairwise properties of components. To find the independent descriptors among 125 descriptors, the feature values are normalized by min-max scaling and then analyzed using pair-wise Pearson correlation and normalized mutual information coefficients\cite{aricode} within the RSTUDIO environment\cite{RSTUDIO}. \\

\noindent \textbf{Machine learning.} We employed the eSVM models for multi-class classification learning tasks.\cite{SVM} The eSVM algorithm comprises of multiple SVM models generated by the bootstrap sampling method \cite{Bootstrap_resampling}. We used the nonlinear Gaussian radial basis function kernel, as implemented in the e1071 package\cite{e1071}. One can generate a large number of training sets using the bootstrap sampling, where samples are randomly drawn with replacement. Every resampling produces two types of samples: (1) in-bag and (2) out-of-bag (OOB), which are used for training and testing the ML models, respectively. The optimization of eSVM hyperparameters is done by the OOB evaluation using grid search. \\ 

\noindent \textbf{Breakdown and Ceteris Paribus methods.} To interpret the trained eSVM model, the BD and CP profile methods as implemented in the DALEX package \cite{Dalexpackage} were applied to compute the contributions of features and individual profiles to ML prediction, respectively. The $k$-mean clustering algorithm from the factoextra package \cite{kmeans_package} was used to divide the dataset containing the BD values into clusters in an unsupervised fashion. 
Local feature importance is analyzed based on the averaged BD data by identifying the correlation between each cluster and the phase selections as predicted by the BD method. The global variable importance of the eSVM is obtained by averaging the outputs of global variable importance for each individual SVM part across all the bootstrap samples. \\ 

\noindent \textbf{Web Application.} Applications developed with the Shiny package\cite{RShiny} in the R programming language allow users to interactively engage with models defined in the server end (server.R). The front end of the application, contained in the user-interface script (ui.R), takes a user inputted string composed of element symbols followed by the amount of the element (e.g., Al1.0V1.0Nb1.0T1.0) representing the composition of the high entropy alloy. The trained eSVM model in the backend generates the phase probability for the given composition. Additionally, the users can obtain the set of 12 descriptors (Table 1), generated using an R script based on the Magpie package. For each new composition, the user can add the phase probability and descriptor information to a dynamic history, able to be exported as a comma separated value file at the end of the session. For each of the 1,367 points in the training set, users can see the associated BD plot and CP profiles. The web app can be accessed at \url{https://adaptivedesign.shinyapps.io/AIRHEAD/}.

\section*{Data Availability}
The dataset used for the ML study is freely available in our Web App (\url{https://adaptivedesign.shinyapps.io/AIRHEAD/}) and on Figshare\cite{FigshareDataset}.

\section*{Competing Interests}
The authors declare that there are no competing interests.

\section*{Author Contribution}
The study was planned by K.L, T.Q.H and P.V.B. The manuscript was prepared by K.L, M.V.A, P.D, T.Q.H and P.V.B. The data set construction was done by K.L and T.Q.H. The machine learning studies were performed by K.L and M.V.A. The web app was built by P.D, K.L, and P.V.B. All authors discussed the results, wrote, and commented on the manuscript.

\begin{acknowledgments}
Research was sponsored by the Defense Advanced Research Project Agency (DARPA) and The Army Research Office and was accomplished under Grant Number W911NF-20-1-0289. The views and conclusions contained in this document are those of the authors and should not be interpreted as representing the official policies, either expressed or implied, of DARPA, the Army Research Office, or the U.S. Government. The U.S. Government is authorized to reproduce and distribute reprints for Government purposes notwithstanding any copyright notation herein.
\end{acknowledgments}



\begin{thebibliography}{76}%
\makeatletter
\providecommand \@ifxundefined [1]{%
 \@ifx{#1\undefined}
}%
\providecommand \@ifnum [1]{%
 \ifnum #1\expandafter \@firstoftwo
 \else \expandafter \@secondoftwo
 \fi
}%
\providecommand \@ifx [1]{%
 \ifx #1\expandafter \@firstoftwo
 \else \expandafter \@secondoftwo
 \fi
}%
\providecommand \natexlab [1]{#1}%
\providecommand \enquote  [1]{``#1''}%
\providecommand \bibnamefont  [1]{#1}%
\providecommand \bibfnamefont [1]{#1}%
\providecommand \citenamefont [1]{#1}%
\providecommand \href@noop [0]{\@secondoftwo}%
\providecommand \href [0]{\begingroup \@sanitize@url \@href}%
\providecommand \@href[1]{\@@startlink{#1}\@@href}%
\providecommand \@@href[1]{\endgroup#1\@@endlink}%
\providecommand \@sanitize@url [0]{\catcode `\\12\catcode `\$12\catcode
  `\&12\catcode `\#12\catcode `\^12\catcode `\_12\catcode `\%12\relax}%
\providecommand \@@startlink[1]{}%
\providecommand \@@endlink[0]{}%
\providecommand \url  [0]{\begingroup\@sanitize@url \@url }%
\providecommand \@url [1]{\endgroup\@href {#1}{\urlprefix }}%
\providecommand \urlprefix  [0]{URL }%
\providecommand \Eprint [0]{\href }%
\providecommand \doibase [0]{http://dx.doi.org/}%
\providecommand \selectlanguage [0]{\@gobble}%
\providecommand \bibinfo  [0]{\@secondoftwo}%
\providecommand \bibfield  [0]{\@secondoftwo}%
\providecommand \translation [1]{[#1]}%
\providecommand \BibitemOpen [0]{}%
\providecommand \bibitemStop [0]{}%
\providecommand \bibitemNoStop [0]{.\EOS\space}%
\providecommand \EOS [0]{\spacefactor3000\relax}%
\providecommand \BibitemShut  [1]{\csname bibitem#1\endcsname}%
\let\auto@bib@innerbib\@empty
\bibitem [{\citenamefont {Senkov}\ \emph {et~al.}(2015)\citenamefont {Senkov},
  \citenamefont {Miller}, \citenamefont {Miracle},\ and\ \citenamefont
  {Woodward}}]{senkov2015accelerated}%
  \BibitemOpen
  \bibfield  {author} {\bibinfo {author} {\bibfnamefont {O.}~\bibnamefont
  {Senkov}}, \bibinfo {author} {\bibfnamefont {J.}~\bibnamefont {Miller}},
  \bibinfo {author} {\bibfnamefont {D.}~\bibnamefont {Miracle}}, \ and\
  \bibinfo {author} {\bibfnamefont {C.}~\bibnamefont {Woodward}},\ }\href@noop
  {} {\bibfield  {journal} {\bibinfo  {journal} {Nature Communications}\
  }\textbf {\bibinfo {volume} {6}},\ \bibinfo {pages} {1} (\bibinfo {year}
  {2015})}\BibitemShut {NoStop}%
\bibitem [{\citenamefont {Cantor}\ \emph {et~al.}(2004)\citenamefont {Cantor},
  \citenamefont {Chang}, \citenamefont {Knight},\ and\ \citenamefont
  {Vincent}}]{CANTOR2004213}%
  \BibitemOpen
  \bibfield  {author} {\bibinfo {author} {\bibfnamefont {B.}~\bibnamefont
  {Cantor}}, \bibinfo {author} {\bibfnamefont {I.}~\bibnamefont {Chang}},
  \bibinfo {author} {\bibfnamefont {P.}~\bibnamefont {Knight}}, \ and\ \bibinfo
  {author} {\bibfnamefont {A.}~\bibnamefont {Vincent}},\ }\href {\doibase
  https://doi.org/10.1016/j.msea.2003.10.257} {\bibfield  {journal} {\bibinfo
  {journal} {Materials Science and Engineering: A}\ }\textbf {\bibinfo {volume}
  {375-377}},\ \bibinfo {pages} {213} (\bibinfo {year} {2004})}\BibitemShut
  {NoStop}%
\bibitem [{\citenamefont {Yeh}\ \emph {et~al.}(2004)\citenamefont {Yeh},
  \citenamefont {Chen}, \citenamefont {Lin}, \citenamefont {Gan}, \citenamefont
  {Chin}, \citenamefont {Shun}, \citenamefont {Tsau},\ and\ \citenamefont
  {Chang}}]{HEA_2004}%
  \BibitemOpen
  \bibfield  {author} {\bibinfo {author} {\bibfnamefont {J.-W.}\ \bibnamefont
  {Yeh}}, \bibinfo {author} {\bibfnamefont {S.-K.}\ \bibnamefont {Chen}},
  \bibinfo {author} {\bibfnamefont {S.-J.}\ \bibnamefont {Lin}}, \bibinfo
  {author} {\bibfnamefont {J.-Y.}\ \bibnamefont {Gan}}, \bibinfo {author}
  {\bibfnamefont {T.-S.}\ \bibnamefont {Chin}}, \bibinfo {author}
  {\bibfnamefont {T.-T.}\ \bibnamefont {Shun}}, \bibinfo {author}
  {\bibfnamefont {C.-H.}\ \bibnamefont {Tsau}}, \ and\ \bibinfo {author}
  {\bibfnamefont {S.-Y.}\ \bibnamefont {Chang}},\ }\href {\doibase
  10.1002/adem.200300567} {\bibfield  {journal} {\bibinfo  {journal} {Advanced
  Engineering Materials}\ }\textbf {\bibinfo {volume} {6}},\ \bibinfo {pages}
  {299} (\bibinfo {year} {2004})}\BibitemShut {NoStop}%
\bibitem [{\citenamefont {Zhang}\ \emph {et~al.}(2014)\citenamefont {Zhang},
  \citenamefont {Zuo}, \citenamefont {Tang}, \citenamefont {Gao}, \citenamefont
  {Dahmen}, \citenamefont {Liaw},\ and\ \citenamefont {Lu}}]{ZHANG20141}%
  \BibitemOpen
  \bibfield  {author} {\bibinfo {author} {\bibfnamefont {Y.}~\bibnamefont
  {Zhang}}, \bibinfo {author} {\bibfnamefont {T.~T.}\ \bibnamefont {Zuo}},
  \bibinfo {author} {\bibfnamefont {Z.}~\bibnamefont {Tang}}, \bibinfo {author}
  {\bibfnamefont {M.~C.}\ \bibnamefont {Gao}}, \bibinfo {author} {\bibfnamefont
  {K.~A.}\ \bibnamefont {Dahmen}}, \bibinfo {author} {\bibfnamefont {P.~K.}\
  \bibnamefont {Liaw}}, \ and\ \bibinfo {author} {\bibfnamefont {Z.~P.}\
  \bibnamefont {Lu}},\ }\href {\doibase
  https://doi.org/10.1016/j.pmatsci.2013.10.001} {\bibfield  {journal}
  {\bibinfo  {journal} {Progress in Materials Science}\ }\textbf {\bibinfo
  {volume} {61}},\ \bibinfo {pages} {1} (\bibinfo {year} {2014})}\BibitemShut
  {NoStop}%
\bibitem [{\citenamefont {Senkov}\ \emph {et~al.}(2018)\citenamefont {Senkov},
  \citenamefont {Miracle}, \citenamefont {Chaput},\ and\ \citenamefont
  {Couzinie}}]{senkov_miracle_chaput_couzinie_2018}%
  \BibitemOpen
  \bibfield  {author} {\bibinfo {author} {\bibfnamefont {O.~N.}\ \bibnamefont
  {Senkov}}, \bibinfo {author} {\bibfnamefont {D.~B.}\ \bibnamefont {Miracle}},
  \bibinfo {author} {\bibfnamefont {K.~J.}\ \bibnamefont {Chaput}}, \ and\
  \bibinfo {author} {\bibfnamefont {J.-P.}\ \bibnamefont {Couzinie}},\ }\href
  {\doibase 10.1557/jmr.2018.153} {\bibfield  {journal} {\bibinfo  {journal}
  {Journal of Materials Research}\ }\textbf {\bibinfo {volume} {33}},\ \bibinfo
  {pages} {3092–3128} (\bibinfo {year} {2018})}\BibitemShut {NoStop}%
\bibitem [{\citenamefont {Kumar}\ and\ \citenamefont
  {Gupta}(2016)}]{met6090199}%
  \BibitemOpen
  \bibfield  {author} {\bibinfo {author} {\bibfnamefont {A.}~\bibnamefont
  {Kumar}}\ and\ \bibinfo {author} {\bibfnamefont {M.}~\bibnamefont {Gupta}},\
  }\href@noop {} {\bibfield  {journal} {\bibinfo  {journal} {Metals}\ }\textbf
  {\bibinfo {volume} {6}},\ \bibinfo {pages} {199} (\bibinfo {year}
  {2016})}\BibitemShut {NoStop}%
\bibitem [{\citenamefont {Gandy}\ \emph {et~al.}(2019)\citenamefont {Gandy},
  \citenamefont {Jim}, \citenamefont {Coe}, \citenamefont {Patel},
  \citenamefont {Hardwick}, \citenamefont {Akhmadaliev}, \citenamefont
  {Reeves-McLaren},\ and\ \citenamefont {Goodall}}]{Si_HEA_Radiation}%
  \BibitemOpen
  \bibfield  {author} {\bibinfo {author} {\bibfnamefont {A.~S.}\ \bibnamefont
  {Gandy}}, \bibinfo {author} {\bibfnamefont {B.}~\bibnamefont {Jim}}, \bibinfo
  {author} {\bibfnamefont {G.}~\bibnamefont {Coe}}, \bibinfo {author}
  {\bibfnamefont {D.}~\bibnamefont {Patel}}, \bibinfo {author} {\bibfnamefont
  {L.}~\bibnamefont {Hardwick}}, \bibinfo {author} {\bibfnamefont
  {S.}~\bibnamefont {Akhmadaliev}}, \bibinfo {author} {\bibfnamefont
  {N.}~\bibnamefont {Reeves-McLaren}}, \ and\ \bibinfo {author} {\bibfnamefont
  {R.}~\bibnamefont {Goodall}},\ }\href {\doibase 10.3389/fmats.2019.00146}
  {\bibfield  {journal} {\bibinfo  {journal} {Frontiers in Materials}\ }\textbf
  {\bibinfo {volume} {6}},\ \bibinfo {pages} {146} (\bibinfo {year}
  {2019})}\BibitemShut {NoStop}%
\bibitem [{\citenamefont {Chen}\ \emph
  {et~al.}(2018{\natexlab{a}})\citenamefont {Chen}, \citenamefont {Zhou},
  \citenamefont {Wang}, \citenamefont {Liu}, \citenamefont {Lv}, \citenamefont
  {Yang}, \citenamefont {Xu},\ and\ \citenamefont {Liu}}]{CHEN201815}%
  \BibitemOpen
  \bibfield  {author} {\bibinfo {author} {\bibfnamefont {J.}~\bibnamefont
  {Chen}}, \bibinfo {author} {\bibfnamefont {X.}~\bibnamefont {Zhou}}, \bibinfo
  {author} {\bibfnamefont {W.}~\bibnamefont {Wang}}, \bibinfo {author}
  {\bibfnamefont {B.}~\bibnamefont {Liu}}, \bibinfo {author} {\bibfnamefont
  {Y.}~\bibnamefont {Lv}}, \bibinfo {author} {\bibfnamefont {W.}~\bibnamefont
  {Yang}}, \bibinfo {author} {\bibfnamefont {D.}~\bibnamefont {Xu}}, \ and\
  \bibinfo {author} {\bibfnamefont {Y.}~\bibnamefont {Liu}},\ }\href {\doibase
  https://doi.org/10.1016/j.jallcom.2018.05.067} {\bibfield  {journal}
  {\bibinfo  {journal} {Journal of Alloys and Compounds}\ }\textbf {\bibinfo
  {volume} {760}},\ \bibinfo {pages} {15} (\bibinfo {year}
  {2018}{\natexlab{a}})}\BibitemShut {NoStop}%
\bibitem [{\citenamefont {Miracle}\ \emph {et~al.}(2014)\citenamefont
  {Miracle}, \citenamefont {Miller}, \citenamefont {Senkov}, \citenamefont
  {Woodward}, \citenamefont {Uchic},\ and\ \citenamefont
  {Tiley}}]{Miracle_Review_2014}%
  \BibitemOpen
  \bibfield  {author} {\bibinfo {author} {\bibfnamefont {D.~B.}\ \bibnamefont
  {Miracle}}, \bibinfo {author} {\bibfnamefont {J.~D.}\ \bibnamefont {Miller}},
  \bibinfo {author} {\bibfnamefont {O.~N.}\ \bibnamefont {Senkov}}, \bibinfo
  {author} {\bibfnamefont {C.}~\bibnamefont {Woodward}}, \bibinfo {author}
  {\bibfnamefont {M.~D.}\ \bibnamefont {Uchic}}, \ and\ \bibinfo {author}
  {\bibfnamefont {J.}~\bibnamefont {Tiley}},\ }\href {\doibase
  10.3390/e16010494} {\bibfield  {journal} {\bibinfo  {journal} {Entropy}\
  }\textbf {\bibinfo {volume} {16}},\ \bibinfo {pages} {494} (\bibinfo {year}
  {2014})}\BibitemShut {NoStop}%
\bibitem [{\citenamefont {Praveen}\ and\ \citenamefont
  {Kim}(2018)}]{doi:10.1002/adem.201700645}%
  \BibitemOpen
  \bibfield  {author} {\bibinfo {author} {\bibfnamefont {S.}~\bibnamefont
  {Praveen}}\ and\ \bibinfo {author} {\bibfnamefont {H.~S.}\ \bibnamefont
  {Kim}},\ }\href {\doibase 10.1002/adem.201700645} {\bibfield  {journal}
  {\bibinfo  {journal} {Advanced Engineering Materials}\ }\textbf {\bibinfo
  {volume} {20}},\ \bibinfo {pages} {1700645} (\bibinfo {year}
  {2018})}\BibitemShut {NoStop}%
\bibitem [{\citenamefont {Miracle}(2019)}]{Miracle_2019_review}%
  \BibitemOpen
  \bibfield  {author} {\bibinfo {author} {\bibfnamefont {D.}~\bibnamefont
  {Miracle}},\ }\href@noop {} {\bibfield  {journal} {\bibinfo  {journal}
  {Nature Communications}\ }\textbf {\bibinfo {volume} {10}},\ \bibinfo {pages}
  {1805} (\bibinfo {year} {2019})}\BibitemShut {NoStop}%
\bibitem [{\citenamefont {George}, \citenamefont {Raabe},\ and\ \citenamefont
  {Ritchie}(2019)}]{HEA_NRM_2019}%
  \BibitemOpen
  \bibfield  {author} {\bibinfo {author} {\bibfnamefont {E.~P.}\ \bibnamefont
  {George}}, \bibinfo {author} {\bibfnamefont {D.}~\bibnamefont {Raabe}}, \
  and\ \bibinfo {author} {\bibfnamefont {R.~O.}\ \bibnamefont {Ritchie}},\
  }\href@noop {} {\bibfield  {journal} {\bibinfo  {journal} {Nature Reviews
  Materials}\ }\textbf {\bibinfo {volume} {4}},\ \bibinfo {pages} {515}
  (\bibinfo {year} {2019})}\BibitemShut {NoStop}%
\bibitem [{\citenamefont {Oses}, \citenamefont {Toher},\ and\ \citenamefont
  {Curtarolo}(2020)}]{HE_Ceramics}%
  \BibitemOpen
  \bibfield  {author} {\bibinfo {author} {\bibfnamefont {C.}~\bibnamefont
  {Oses}}, \bibinfo {author} {\bibfnamefont {C.}~\bibnamefont {Toher}}, \ and\
  \bibinfo {author} {\bibfnamefont {S.}~\bibnamefont {Curtarolo}},\ }\href@noop
  {} {\bibfield  {journal} {\bibinfo  {journal} {Nature Reviews Materials}\
  }\textbf {\bibinfo {volume} {5}},\ \bibinfo {pages} {295} (\bibinfo {year}
  {2020})}\BibitemShut {NoStop}%
\bibitem [{\citenamefont {Zhou}\ \emph
  {et~al.}(2019{\natexlab{a}})\citenamefont {Zhou}, \citenamefont {Jiang},
  \citenamefont {Huang}, \citenamefont {Qin}, \citenamefont {Hu},\ and\
  \citenamefont {Luo}}]{zhou2019single}%
  \BibitemOpen
  \bibfield  {author} {\bibinfo {author} {\bibfnamefont {N.}~\bibnamefont
  {Zhou}}, \bibinfo {author} {\bibfnamefont {S.}~\bibnamefont {Jiang}},
  \bibinfo {author} {\bibfnamefont {T.}~\bibnamefont {Huang}}, \bibinfo
  {author} {\bibfnamefont {M.}~\bibnamefont {Qin}}, \bibinfo {author}
  {\bibfnamefont {T.}~\bibnamefont {Hu}}, \ and\ \bibinfo {author}
  {\bibfnamefont {J.}~\bibnamefont {Luo}},\ }\href@noop {} {\bibfield
  {journal} {\bibinfo  {journal} {Science Bulletin}\ }\textbf {\bibinfo
  {volume} {64}},\ \bibinfo {pages} {856} (\bibinfo {year}
  {2019}{\natexlab{a}})}\BibitemShut {NoStop}%
\bibitem [{\citenamefont {Wong}\ \emph {et~al.}(2018)\citenamefont {Wong},
  \citenamefont {Shun}, \citenamefont {Chang},\ and\ \citenamefont
  {Lee}}]{WONG2018146}%
  \BibitemOpen
  \bibfield  {author} {\bibinfo {author} {\bibfnamefont {S.-K.}\ \bibnamefont
  {Wong}}, \bibinfo {author} {\bibfnamefont {T.-T.}\ \bibnamefont {Shun}},
  \bibinfo {author} {\bibfnamefont {C.-H.}\ \bibnamefont {Chang}}, \ and\
  \bibinfo {author} {\bibfnamefont {C.-F.}\ \bibnamefont {Lee}},\ }\href
  {\doibase https://doi.org/10.1016/j.matchemphys.2017.07.085} {\bibfield
  {journal} {\bibinfo  {journal} {Materials Chemistry and Physics}\ }\textbf
  {\bibinfo {volume} {210}},\ \bibinfo {pages} {146} (\bibinfo {year}
  {2018})},\ \bibinfo {note} {high-Entropy Materials}\BibitemShut {NoStop}%
\bibitem [{\citenamefont {Li}\ \emph {et~al.}(2016)\citenamefont {Li},
  \citenamefont {Pradeep}, \citenamefont {Deng}, \citenamefont {Raabe},\ and\
  \citenamefont {Tasan}}]{li_metastable_2016}%
  \BibitemOpen
  \bibfield  {author} {\bibinfo {author} {\bibfnamefont {Z.}~\bibnamefont
  {Li}}, \bibinfo {author} {\bibfnamefont {K.~G.}\ \bibnamefont {Pradeep}},
  \bibinfo {author} {\bibfnamefont {Y.}~\bibnamefont {Deng}}, \bibinfo {author}
  {\bibfnamefont {D.}~\bibnamefont {Raabe}}, \ and\ \bibinfo {author}
  {\bibfnamefont {C.~C.}\ \bibnamefont {Tasan}},\ }\href {\doibase
  10.1038/nature17981} {\bibfield  {journal} {\bibinfo  {journal} {Nature}\
  }\textbf {\bibinfo {volume} {534}},\ \bibinfo {pages} {227} (\bibinfo {year}
  {2016})}\BibitemShut {NoStop}%
\bibitem [{\citenamefont {Chen}\ \emph
  {et~al.}(2018{\natexlab{b}})\citenamefont {Chen}, \citenamefont {Qin},
  \citenamefont {Zheng}, \citenamefont {Wang}, \citenamefont {Su},
  \citenamefont {Chiu}, \citenamefont {Ding}, \citenamefont {Guo},\ and\
  \citenamefont {Fu}}]{CHEN2018129}%
  \BibitemOpen
  \bibfield  {author} {\bibinfo {author} {\bibfnamefont {R.}~\bibnamefont
  {Chen}}, \bibinfo {author} {\bibfnamefont {G.}~\bibnamefont {Qin}}, \bibinfo
  {author} {\bibfnamefont {H.}~\bibnamefont {Zheng}}, \bibinfo {author}
  {\bibfnamefont {L.}~\bibnamefont {Wang}}, \bibinfo {author} {\bibfnamefont
  {Y.}~\bibnamefont {Su}}, \bibinfo {author} {\bibfnamefont {Y.}~\bibnamefont
  {Chiu}}, \bibinfo {author} {\bibfnamefont {H.}~\bibnamefont {Ding}}, \bibinfo
  {author} {\bibfnamefont {J.}~\bibnamefont {Guo}}, \ and\ \bibinfo {author}
  {\bibfnamefont {H.}~\bibnamefont {Fu}},\ }\href {\doibase
  https://doi.org/10.1016/j.actamat.2017.10.058} {\bibfield  {journal}
  {\bibinfo  {journal} {Acta Materialia}\ }\textbf {\bibinfo {volume} {144}},\
  \bibinfo {pages} {129} (\bibinfo {year} {2018}{\natexlab{b}})}\BibitemShut
  {NoStop}%
\bibitem [{\citenamefont {Tang}\ \emph {et~al.}(2019)\citenamefont {Tang},
  \citenamefont {Zhang}, \citenamefont {Cai}, \citenamefont {Zhou},\ and\
  \citenamefont {Wang}}]{tang2019designing}%
  \BibitemOpen
  \bibfield  {author} {\bibinfo {author} {\bibfnamefont {Z.}~\bibnamefont
  {Tang}}, \bibinfo {author} {\bibfnamefont {S.}~\bibnamefont {Zhang}},
  \bibinfo {author} {\bibfnamefont {R.}~\bibnamefont {Cai}}, \bibinfo {author}
  {\bibfnamefont {Q.}~\bibnamefont {Zhou}}, \ and\ \bibinfo {author}
  {\bibfnamefont {H.}~\bibnamefont {Wang}},\ }\href@noop {} {\bibfield
  {journal} {\bibinfo  {journal} {Metallurgical and Materials Transactions A}\
  }\textbf {\bibinfo {volume} {50}},\ \bibinfo {pages} {1888} (\bibinfo {year}
  {2019})}\BibitemShut {NoStop}%
\bibitem [{\citenamefont {Feuerbacher}, \citenamefont {Lienig},\ and\
  \citenamefont {Thomas}(2018)}]{feuerbacher2018single}%
  \BibitemOpen
  \bibfield  {author} {\bibinfo {author} {\bibfnamefont {M.}~\bibnamefont
  {Feuerbacher}}, \bibinfo {author} {\bibfnamefont {T.}~\bibnamefont {Lienig}},
  \ and\ \bibinfo {author} {\bibfnamefont {C.}~\bibnamefont {Thomas}},\
  }\href@noop {} {\bibfield  {journal} {\bibinfo  {journal} {Scripta
  Materialia}\ }\textbf {\bibinfo {volume} {152}},\ \bibinfo {pages} {40}
  (\bibinfo {year} {2018})}\BibitemShut {NoStop}%
\bibitem [{\citenamefont {Zhang}\ and\ \citenamefont
  {Gao}(2016)}]{zhang2016calphad}%
  \BibitemOpen
  \bibfield  {author} {\bibinfo {author} {\bibfnamefont {C.}~\bibnamefont
  {Zhang}}\ and\ \bibinfo {author} {\bibfnamefont {M.~C.}\ \bibnamefont
  {Gao}},\ }in\ \href@noop {} {\emph {\bibinfo {booktitle} {{High-Entropy
  Alloys}}}}\ (\bibinfo  {publisher} {Springer},\ \bibinfo {year} {2016})\ pp.\
  \bibinfo {pages} {399--444}\BibitemShut {NoStop}%
\bibitem [{\citenamefont {Feng}\ \emph {et~al.}(2021)\citenamefont {Feng},
  \citenamefont {Zhang}, \citenamefont {Gao}, \citenamefont {Pei},
  \citenamefont {Zhang}, \citenamefont {Chen}, \citenamefont {Ma},
  \citenamefont {An}, \citenamefont {Poplawsky}, \citenamefont {Ouyang} \emph
  {et~al.}}]{feng2021high}%
  \BibitemOpen
  \bibfield  {author} {\bibinfo {author} {\bibfnamefont {R.}~\bibnamefont
  {Feng}}, \bibinfo {author} {\bibfnamefont {C.}~\bibnamefont {Zhang}},
  \bibinfo {author} {\bibfnamefont {M.~C.}\ \bibnamefont {Gao}}, \bibinfo
  {author} {\bibfnamefont {Z.}~\bibnamefont {Pei}}, \bibinfo {author}
  {\bibfnamefont {F.}~\bibnamefont {Zhang}}, \bibinfo {author} {\bibfnamefont
  {Y.}~\bibnamefont {Chen}}, \bibinfo {author} {\bibfnamefont {D.}~\bibnamefont
  {Ma}}, \bibinfo {author} {\bibfnamefont {K.}~\bibnamefont {An}}, \bibinfo
  {author} {\bibfnamefont {J.~D.}\ \bibnamefont {Poplawsky}}, \bibinfo {author}
  {\bibfnamefont {L.}~\bibnamefont {Ouyang}},  \emph {et~al.},\ }\href@noop {}
  {\bibfield  {journal} {\bibinfo  {journal} {Nature Communications}\ }\textbf
  {\bibinfo {volume} {12}},\ \bibinfo {pages} {1} (\bibinfo {year}
  {2021})}\BibitemShut {NoStop}%
\bibitem [{\citenamefont {Qi}, \citenamefont {Cheung},\ and\ \citenamefont
  {Poon}(2019)}]{Poon_HEA_SciRep}%
  \BibitemOpen
  \bibfield  {author} {\bibinfo {author} {\bibfnamefont {J.}~\bibnamefont
  {Qi}}, \bibinfo {author} {\bibfnamefont {A.~M.}\ \bibnamefont {Cheung}}, \
  and\ \bibinfo {author} {\bibfnamefont {S.~J.}\ \bibnamefont {Poon}},\
  }\href@noop {} {\bibfield  {journal} {\bibinfo  {journal} {Scientific
  Reports}\ }\textbf {\bibinfo {volume} {9}},\ \bibinfo {pages} {15501}
  (\bibinfo {year} {2019})}\BibitemShut {NoStop}%
\bibitem [{\citenamefont {Islam}, \citenamefont {Huang},\ and\ \citenamefont
  {Zhuang}(2018)}]{ISLAM2018230}%
  \BibitemOpen
  \bibfield  {author} {\bibinfo {author} {\bibfnamefont {N.}~\bibnamefont
  {Islam}}, \bibinfo {author} {\bibfnamefont {W.}~\bibnamefont {Huang}}, \ and\
  \bibinfo {author} {\bibfnamefont {H.~L.}\ \bibnamefont {Zhuang}},\ }\href
  {\doibase https://doi.org/10.1016/j.commatsci.2018.04.003} {\bibfield
  {journal} {\bibinfo  {journal} {Computational Materials Science}\ }\textbf
  {\bibinfo {volume} {150}},\ \bibinfo {pages} {230} (\bibinfo {year}
  {2018})}\BibitemShut {NoStop}%
\bibitem [{\citenamefont {Kim}\ \emph {et~al.}(2019)\citenamefont {Kim},
  \citenamefont {Diao}, \citenamefont {Lee}, \citenamefont {Samaei},
  \citenamefont {Phan}, \citenamefont {{de Jong}}, \citenamefont {An},
  \citenamefont {Ma}, \citenamefont {Liaw},\ and\ \citenamefont
  {Chen}}]{KIM2019124}%
  \BibitemOpen
  \bibfield  {author} {\bibinfo {author} {\bibfnamefont {G.}~\bibnamefont
  {Kim}}, \bibinfo {author} {\bibfnamefont {H.}~\bibnamefont {Diao}}, \bibinfo
  {author} {\bibfnamefont {C.}~\bibnamefont {Lee}}, \bibinfo {author}
  {\bibfnamefont {A.}~\bibnamefont {Samaei}}, \bibinfo {author} {\bibfnamefont
  {T.}~\bibnamefont {Phan}}, \bibinfo {author} {\bibfnamefont {M.}~\bibnamefont
  {{de Jong}}}, \bibinfo {author} {\bibfnamefont {K.}~\bibnamefont {An}},
  \bibinfo {author} {\bibfnamefont {D.}~\bibnamefont {Ma}}, \bibinfo {author}
  {\bibfnamefont {P.~K.}\ \bibnamefont {Liaw}}, \ and\ \bibinfo {author}
  {\bibfnamefont {W.}~\bibnamefont {Chen}},\ }\href {\doibase
  https://doi.org/10.1016/j.actamat.2019.09.026} {\bibfield  {journal}
  {\bibinfo  {journal} {Acta Materialia}\ }\textbf {\bibinfo {volume} {181}},\
  \bibinfo {pages} {124} (\bibinfo {year} {2019})}\BibitemShut {NoStop}%
\bibitem [{\citenamefont {Zhou}\ \emph
  {et~al.}(2019{\natexlab{b}})\citenamefont {Zhou}, \citenamefont {Zhou},
  \citenamefont {He}, \citenamefont {Ding}, \citenamefont {Li},\ and\
  \citenamefont {Yang}}]{zhou_machine_2019}%
  \BibitemOpen
  \bibfield  {author} {\bibinfo {author} {\bibfnamefont {Z.}~\bibnamefont
  {Zhou}}, \bibinfo {author} {\bibfnamefont {Y.}~\bibnamefont {Zhou}}, \bibinfo
  {author} {\bibfnamefont {Q.}~\bibnamefont {He}}, \bibinfo {author}
  {\bibfnamefont {Z.}~\bibnamefont {Ding}}, \bibinfo {author} {\bibfnamefont
  {F.}~\bibnamefont {Li}}, \ and\ \bibinfo {author} {\bibfnamefont
  {Y.}~\bibnamefont {Yang}},\ }\href {\doibase 10.1038/s41524-019-0265-1}
  {\bibfield  {journal} {\bibinfo  {journal} {npj Computational Materials}\
  }\textbf {\bibinfo {volume} {5}},\ \bibinfo {pages} {128} (\bibinfo {year}
  {2019}{\natexlab{b}})}\BibitemShut {NoStop}%
\bibitem [{\citenamefont {Huang}, \citenamefont {Martin},\ and\ \citenamefont
  {Zhuang}(2019)}]{HUANG2019225}%
  \BibitemOpen
  \bibfield  {author} {\bibinfo {author} {\bibfnamefont {W.}~\bibnamefont
  {Huang}}, \bibinfo {author} {\bibfnamefont {P.}~\bibnamefont {Martin}}, \
  and\ \bibinfo {author} {\bibfnamefont {H.~L.}\ \bibnamefont {Zhuang}},\
  }\href {\doibase https://doi.org/10.1016/j.actamat.2019.03.012} {\bibfield
  {journal} {\bibinfo  {journal} {Acta Materialia}\ }\textbf {\bibinfo {volume}
  {169}},\ \bibinfo {pages} {225} (\bibinfo {year} {2019})}\BibitemShut
  {NoStop}%
\bibitem [{\citenamefont {Li}\ and\ \citenamefont
  {Guo}(2019)}]{PhysRevMaterials.3.095005}%
  \BibitemOpen
  \bibfield  {author} {\bibinfo {author} {\bibfnamefont {Y.}~\bibnamefont
  {Li}}\ and\ \bibinfo {author} {\bibfnamefont {W.}~\bibnamefont {Guo}},\
  }\href {\doibase 10.1103/PhysRevMaterials.3.095005} {\bibfield  {journal}
  {\bibinfo  {journal} {Phys. Rev. Materials}\ }\textbf {\bibinfo {volume}
  {3}},\ \bibinfo {pages} {095005} (\bibinfo {year} {2019})}\BibitemShut
  {NoStop}%
\bibitem [{\citenamefont {Qu}\ \emph {et~al.}(2019)\citenamefont {Qu},
  \citenamefont {Chen}, \citenamefont {Lai}, \citenamefont {Liu},\ and\
  \citenamefont {Zhu}}]{QU2019299}%
  \BibitemOpen
  \bibfield  {author} {\bibinfo {author} {\bibfnamefont {N.}~\bibnamefont
  {Qu}}, \bibinfo {author} {\bibfnamefont {Y.}~\bibnamefont {Chen}}, \bibinfo
  {author} {\bibfnamefont {Z.}~\bibnamefont {Lai}}, \bibinfo {author}
  {\bibfnamefont {Y.}~\bibnamefont {Liu}}, \ and\ \bibinfo {author}
  {\bibfnamefont {J.}~\bibnamefont {Zhu}},\ }\href {\doibase
  https://doi.org/10.1016/j.promfg.2019.12.051} {\bibfield  {journal} {\bibinfo
   {journal} {Procedia Manufacturing}\ }\textbf {\bibinfo {volume} {37}},\
  \bibinfo {pages} {299} (\bibinfo {year} {2019})},\ \bibinfo {note} {{Physical
  and Numerical Simulation of Materials Processing IX}}\BibitemShut {NoStop}%
\bibitem [{\citenamefont {Kaufmann}\ and\ \citenamefont
  {Vecchio}(2020)}]{KAUFMANN2020178}%
  \BibitemOpen
  \bibfield  {author} {\bibinfo {author} {\bibfnamefont {K.}~\bibnamefont
  {Kaufmann}}\ and\ \bibinfo {author} {\bibfnamefont {K.~S.}\ \bibnamefont
  {Vecchio}},\ }\href {\doibase https://doi.org/10.1016/j.actamat.2020.07.065}
  {\bibfield  {journal} {\bibinfo  {journal} {Acta Materialia}\ }\textbf
  {\bibinfo {volume} {198}},\ \bibinfo {pages} {178} (\bibinfo {year}
  {2020})}\BibitemShut {NoStop}%
\bibitem [{\citenamefont {Zhang}\ \emph
  {et~al.}(2020{\natexlab{a}})\citenamefont {Zhang}, \citenamefont {Chen},
  \citenamefont {Tao}, \citenamefont {Cai}, \citenamefont {Liu}, \citenamefont
  {Ouyang}, \citenamefont {Peng},\ and\ \citenamefont {Du}}]{ZHANG2020108835}%
  \BibitemOpen
  \bibfield  {author} {\bibinfo {author} {\bibfnamefont {L.}~\bibnamefont
  {Zhang}}, \bibinfo {author} {\bibfnamefont {H.}~\bibnamefont {Chen}},
  \bibinfo {author} {\bibfnamefont {X.}~\bibnamefont {Tao}}, \bibinfo {author}
  {\bibfnamefont {H.}~\bibnamefont {Cai}}, \bibinfo {author} {\bibfnamefont
  {J.}~\bibnamefont {Liu}}, \bibinfo {author} {\bibfnamefont {Y.}~\bibnamefont
  {Ouyang}}, \bibinfo {author} {\bibfnamefont {Q.}~\bibnamefont {Peng}}, \ and\
  \bibinfo {author} {\bibfnamefont {Y.}~\bibnamefont {Du}},\ }\href {\doibase
  https://doi.org/10.1016/j.matdes.2020.108835} {\bibfield  {journal} {\bibinfo
   {journal} {Materials \& Design}\ }\textbf {\bibinfo {volume} {193}},\
  \bibinfo {pages} {108835} (\bibinfo {year} {2020}{\natexlab{a}})}\BibitemShut
  {NoStop}%
\bibitem [{\citenamefont {Dai}\ \emph {et~al.}(2020)\citenamefont {Dai},
  \citenamefont {Xu}, \citenamefont {Wei}, \citenamefont {Ding}, \citenamefont
  {Xu}, \citenamefont {Zhang},\ and\ \citenamefont {Zhang}}]{DAI2020109618}%
  \BibitemOpen
  \bibfield  {author} {\bibinfo {author} {\bibfnamefont {D.}~\bibnamefont
  {Dai}}, \bibinfo {author} {\bibfnamefont {T.}~\bibnamefont {Xu}}, \bibinfo
  {author} {\bibfnamefont {X.}~\bibnamefont {Wei}}, \bibinfo {author}
  {\bibfnamefont {G.}~\bibnamefont {Ding}}, \bibinfo {author} {\bibfnamefont
  {Y.}~\bibnamefont {Xu}}, \bibinfo {author} {\bibfnamefont {J.}~\bibnamefont
  {Zhang}}, \ and\ \bibinfo {author} {\bibfnamefont {H.}~\bibnamefont
  {Zhang}},\ }\href {\doibase https://doi.org/10.1016/j.commatsci.2020.109618}
  {\bibfield  {journal} {\bibinfo  {journal} {Computational Materials Science}\
  }\textbf {\bibinfo {volume} {175}},\ \bibinfo {pages} {109618} (\bibinfo
  {year} {2020})}\BibitemShut {NoStop}%
\bibitem [{\citenamefont {Pei}\ \emph {et~al.}(2020)\citenamefont {Pei},
  \citenamefont {Yin}, \citenamefont {Hawk}, \citenamefont {Alman},\ and\
  \citenamefont {Gao}}]{pei_npjCM2020}%
  \BibitemOpen
  \bibfield  {author} {\bibinfo {author} {\bibfnamefont {Z.}~\bibnamefont
  {Pei}}, \bibinfo {author} {\bibfnamefont {J.}~\bibnamefont {Yin}}, \bibinfo
  {author} {\bibfnamefont {J.~A.}\ \bibnamefont {Hawk}}, \bibinfo {author}
  {\bibfnamefont {D.~E.}\ \bibnamefont {Alman}}, \ and\ \bibinfo {author}
  {\bibfnamefont {M.~C.}\ \bibnamefont {Gao}},\ }\href {\doibase
  10.1038/s41524-020-0308-7} {\bibfield  {journal} {\bibinfo  {journal} {npj
  Computational Materials}\ }\textbf {\bibinfo {volume} {6}},\ \bibinfo {pages}
  {50} (\bibinfo {year} {2020})}\BibitemShut {NoStop}%
\bibitem [{\citenamefont {Zhang}\ \emph
  {et~al.}(2020{\natexlab{b}})\citenamefont {Zhang}, \citenamefont {Wen},
  \citenamefont {Wang}, \citenamefont {Antonov}, \citenamefont {Xue},
  \citenamefont {Bai},\ and\ \citenamefont {Su}}]{ZHANG2020528}%
  \BibitemOpen
  \bibfield  {author} {\bibinfo {author} {\bibfnamefont {Y.}~\bibnamefont
  {Zhang}}, \bibinfo {author} {\bibfnamefont {C.}~\bibnamefont {Wen}}, \bibinfo
  {author} {\bibfnamefont {C.}~\bibnamefont {Wang}}, \bibinfo {author}
  {\bibfnamefont {S.}~\bibnamefont {Antonov}}, \bibinfo {author} {\bibfnamefont
  {D.}~\bibnamefont {Xue}}, \bibinfo {author} {\bibfnamefont {Y.}~\bibnamefont
  {Bai}}, \ and\ \bibinfo {author} {\bibfnamefont {Y.}~\bibnamefont {Su}},\
  }\href {\doibase https://doi.org/10.1016/j.actamat.2019.11.067} {\bibfield
  {journal} {\bibinfo  {journal} {Acta Materialia}\ }\textbf {\bibinfo {volume}
  {185}},\ \bibinfo {pages} {528} (\bibinfo {year}
  {2020}{\natexlab{b}})}\BibitemShut {NoStop}%
\bibitem [{\citenamefont {Risal}\ \emph {et~al.}(2021)\citenamefont {Risal},
  \citenamefont {Zhu}, \citenamefont {Guillen},\ and\ \citenamefont
  {Sun}}]{RISAL2021110389}%
  \BibitemOpen
  \bibfield  {author} {\bibinfo {author} {\bibfnamefont {S.}~\bibnamefont
  {Risal}}, \bibinfo {author} {\bibfnamefont {W.}~\bibnamefont {Zhu}}, \bibinfo
  {author} {\bibfnamefont {P.}~\bibnamefont {Guillen}}, \ and\ \bibinfo
  {author} {\bibfnamefont {L.}~\bibnamefont {Sun}},\ }\href {\doibase
  https://doi.org/10.1016/j.commatsci.2021.110389} {\bibfield  {journal}
  {\bibinfo  {journal} {Computational Materials Science}\ }\textbf {\bibinfo
  {volume} {192}},\ \bibinfo {pages} {110389} (\bibinfo {year}
  {2021})}\BibitemShut {NoStop}%
\bibitem [{\citenamefont {Lee}\ \emph {et~al.}(2021{\natexlab{a}})\citenamefont
  {Lee}, \citenamefont {Byeon}, \citenamefont {Kim}, \citenamefont {Jin},\ and\
  \citenamefont {Lee}}]{LEE2021109260}%
  \BibitemOpen
  \bibfield  {author} {\bibinfo {author} {\bibfnamefont {S.~Y.}\ \bibnamefont
  {Lee}}, \bibinfo {author} {\bibfnamefont {S.}~\bibnamefont {Byeon}}, \bibinfo
  {author} {\bibfnamefont {H.~S.}\ \bibnamefont {Kim}}, \bibinfo {author}
  {\bibfnamefont {H.}~\bibnamefont {Jin}}, \ and\ \bibinfo {author}
  {\bibfnamefont {S.}~\bibnamefont {Lee}},\ }\href {\doibase
  https://doi.org/10.1016/j.matdes.2020.109260} {\bibfield  {journal} {\bibinfo
   {journal} {Materials \& Design}\ }\textbf {\bibinfo {volume} {197}},\
  \bibinfo {pages} {109260} (\bibinfo {year} {2021}{\natexlab{a}})}\BibitemShut
  {NoStop}%
\bibitem [{\citenamefont {Beniwal}\ and\ \citenamefont
  {Ray}(2021)}]{BENIWAL2021110647}%
  \BibitemOpen
  \bibfield  {author} {\bibinfo {author} {\bibfnamefont {D.}~\bibnamefont
  {Beniwal}}\ and\ \bibinfo {author} {\bibfnamefont {P.}~\bibnamefont {Ray}},\
  }\href {\doibase https://doi.org/10.1016/j.commatsci.2021.110647} {\bibfield
  {journal} {\bibinfo  {journal} {Computational Materials Science}\ }\textbf
  {\bibinfo {volume} {197}},\ \bibinfo {pages} {110647} (\bibinfo {year}
  {2021})}\BibitemShut {NoStop}%
\bibitem [{\citenamefont {Yan}, \citenamefont {Lu},\ and\ \citenamefont
  {Wang}(2021)}]{YAN2021110723}%
  \BibitemOpen
  \bibfield  {author} {\bibinfo {author} {\bibfnamefont {Y.}~\bibnamefont
  {Yan}}, \bibinfo {author} {\bibfnamefont {D.}~\bibnamefont {Lu}}, \ and\
  \bibinfo {author} {\bibfnamefont {K.}~\bibnamefont {Wang}},\ }\href {\doibase
  https://doi.org/10.1016/j.commatsci.2021.110723} {\bibfield  {journal}
  {\bibinfo  {journal} {Computational Materials Science}\ }\textbf {\bibinfo
  {volume} {199}},\ \bibinfo {pages} {110723} (\bibinfo {year}
  {2021})}\BibitemShut {NoStop}%
\bibitem [{\citenamefont {Staniak}\ and\ \citenamefont
  {Biecek}(2019)}]{Staniak_2019}%
  \BibitemOpen
  \bibfield  {author} {\bibinfo {author} {\bibfnamefont {M.}~\bibnamefont
  {Staniak}}\ and\ \bibinfo {author} {\bibfnamefont {P.}~\bibnamefont
  {Biecek}},\ }\href {\doibase 10.32614/rj-2018-072} {\bibfield  {journal}
  {\bibinfo  {journal} {The R Journal}\ }\textbf {\bibinfo {volume} {10}},\
  \bibinfo {pages} {395} (\bibinfo {year} {2019})}\BibitemShut {NoStop}%
\bibitem [{\citenamefont {Cortes}\ and\ \citenamefont
  {Vapnik}(1995)}]{cortes_support-vector_1995}%
  \BibitemOpen
  \bibfield  {author} {\bibinfo {author} {\bibfnamefont {C.}~\bibnamefont
  {Cortes}}\ and\ \bibinfo {author} {\bibfnamefont {V.}~\bibnamefont
  {Vapnik}},\ }\href {\doibase 10.1007/BF00994018} {\bibfield  {journal}
  {\bibinfo  {journal} {Machine Learning}\ }\textbf {\bibinfo {volume} {20}},\
  \bibinfo {pages} {273} (\bibinfo {year} {1995})}\BibitemShut {NoStop}%
\bibitem [{\citenamefont {Vapnik}(2006)}]{vapnik_estimation_2006}%
  \BibitemOpen
  \bibfield  {author} {\bibinfo {author} {\bibfnamefont {V.~N.}\ \bibnamefont
  {Vapnik}},\ }\href@noop {} {\emph {\bibinfo {title} {{Estimation of
  dependences based on empirical data: {Empirical} inference science: afterword
  of 2006}}}},\ \bibinfo {edition} {2nd}\ ed.,\ Information science and
  statistics\ (\bibinfo  {publisher} {Springer},\ \bibinfo {address} {New York,
  N.Y},\ \bibinfo {year} {2006})\BibitemShut {NoStop}%
\bibitem [{\citenamefont {Yang}\ and\ \citenamefont
  {Zhang}(2012)}]{YANG2012233}%
  \BibitemOpen
  \bibfield  {author} {\bibinfo {author} {\bibfnamefont {X.}~\bibnamefont
  {Yang}}\ and\ \bibinfo {author} {\bibfnamefont {Y.}~\bibnamefont {Zhang}},\
  }\href {\doibase https://doi.org/10.1016/j.matchemphys.2011.11.021}
  {\bibfield  {journal} {\bibinfo  {journal} {Materials Chemistry and Physics}\
  }\textbf {\bibinfo {volume} {132}},\ \bibinfo {pages} {233} (\bibinfo {year}
  {2012})}\BibitemShut {NoStop}%
\bibitem [{\citenamefont {Hu}\ \emph {et~al.}(2017)\citenamefont {Hu},
  \citenamefont {Guo}, \citenamefont {Wang}, \citenamefont {Yan}, \citenamefont
  {Chen}, \citenamefont {Lu}, \citenamefont {Liu}, \citenamefont {Zou},\ and\
  \citenamefont {Zeng}}]{hu_parametric_2017}%
  \BibitemOpen
  \bibfield  {author} {\bibinfo {author} {\bibfnamefont {Q.}~\bibnamefont
  {Hu}}, \bibinfo {author} {\bibfnamefont {S.}~\bibnamefont {Guo}}, \bibinfo
  {author} {\bibfnamefont {J.}~\bibnamefont {Wang}}, \bibinfo {author}
  {\bibfnamefont {Y.}~\bibnamefont {Yan}}, \bibinfo {author} {\bibfnamefont
  {S.}~\bibnamefont {Chen}}, \bibinfo {author} {\bibfnamefont {D.}~\bibnamefont
  {Lu}}, \bibinfo {author} {\bibfnamefont {K.}~\bibnamefont {Liu}}, \bibinfo
  {author} {\bibfnamefont {J.}~\bibnamefont {Zou}}, \ and\ \bibinfo {author}
  {\bibfnamefont {X.}~\bibnamefont {Zeng}},\ }\href {\doibase
  10.1038/srep39917} {\bibfield  {journal} {\bibinfo  {journal} {Scientific
  Reports}\ }\textbf {\bibinfo {volume} {7}},\ \bibinfo {pages} {39917}
  (\bibinfo {year} {2017})}\BibitemShut {NoStop}%
\bibitem [{\citenamefont {Senkov}\ and\ \citenamefont
  {Miracle}(2016)}]{SENKOV2016603}%
  \BibitemOpen
  \bibfield  {author} {\bibinfo {author} {\bibfnamefont {O.}~\bibnamefont
  {Senkov}}\ and\ \bibinfo {author} {\bibfnamefont {D.}~\bibnamefont
  {Miracle}},\ }\href {\doibase https://doi.org/10.1016/j.jallcom.2015.10.279}
  {\bibfield  {journal} {\bibinfo  {journal} {Journal of Alloys and Compounds}\
  }\textbf {\bibinfo {volume} {658}},\ \bibinfo {pages} {603} (\bibinfo {year}
  {2016})}\BibitemShut {NoStop}%
\bibitem [{\citenamefont {Guo}\ \emph {et~al.}(2013)\citenamefont {Guo},
  \citenamefont {Hu}, \citenamefont {Ng},\ and\ \citenamefont
  {Liu}}]{GUO201396}%
  \BibitemOpen
  \bibfield  {author} {\bibinfo {author} {\bibfnamefont {S.}~\bibnamefont
  {Guo}}, \bibinfo {author} {\bibfnamefont {Q.}~\bibnamefont {Hu}}, \bibinfo
  {author} {\bibfnamefont {C.}~\bibnamefont {Ng}}, \ and\ \bibinfo {author}
  {\bibfnamefont {C.}~\bibnamefont {Liu}},\ }\href {\doibase
  https://doi.org/10.1016/j.intermet.2013.05.002} {\bibfield  {journal}
  {\bibinfo  {journal} {Intermetallics}\ }\textbf {\bibinfo {volume} {41}},\
  \bibinfo {pages} {96} (\bibinfo {year} {2013})}\BibitemShut {NoStop}%
\bibitem [{\citenamefont {Toda-Caraballo}\ and\ \citenamefont {del
  Castillo}(2016)}]{TODACARABALLO201676}%
  \BibitemOpen
  \bibfield  {author} {\bibinfo {author} {\bibfnamefont {I.}~\bibnamefont
  {Toda-Caraballo}}\ and\ \bibinfo {author} {\bibfnamefont {P.~R.-D.}\
  \bibnamefont {del Castillo}},\ }\href {\doibase
  https://doi.org/10.1016/j.intermet.2015.12.011} {\bibfield  {journal}
  {\bibinfo  {journal} {Intermetallics}\ }\textbf {\bibinfo {volume} {71}},\
  \bibinfo {pages} {76} (\bibinfo {year} {2016})}\BibitemShut {NoStop}%
\bibitem [{\citenamefont {Miracle}\ and\ \citenamefont
  {Senkov}(2017)}]{Miracle_ActMat2017}%
  \BibitemOpen
  \bibfield  {author} {\bibinfo {author} {\bibfnamefont {D.}~\bibnamefont
  {Miracle}}\ and\ \bibinfo {author} {\bibfnamefont {O.}~\bibnamefont
  {Senkov}},\ }\href {\doibase https://doi.org/10.1016/j.actamat.2016.08.081}
  {\bibfield  {journal} {\bibinfo  {journal} {Acta Materialia}\ }\textbf
  {\bibinfo {volume} {122}},\ \bibinfo {pages} {448} (\bibinfo {year}
  {2017})}\BibitemShut {NoStop}%
\bibitem [{\citenamefont {Parlinski}, \citenamefont {Li},\ and\ \citenamefont
  {Kawazoe}(1997)}]{frozen_phonon}%
  \BibitemOpen
  \bibfield  {author} {\bibinfo {author} {\bibfnamefont {K.}~\bibnamefont
  {Parlinski}}, \bibinfo {author} {\bibfnamefont {Z.~Q.}\ \bibnamefont {Li}}, \
  and\ \bibinfo {author} {\bibfnamefont {Y.}~\bibnamefont {Kawazoe}},\ }\href
  {\doibase 10.1103/PhysRevLett.78.4063} {\bibfield  {journal} {\bibinfo
  {journal} {Phys. Rev. Lett.}\ }\textbf {\bibinfo {volume} {78}},\ \bibinfo
  {pages} {4063} (\bibinfo {year} {1997})}\BibitemShut {NoStop}%
\bibitem [{\citenamefont {Gao}\ \emph {et~al.}(2017)\citenamefont {Gao},
  \citenamefont {Zhang}, \citenamefont {Gao}, \citenamefont {Zhang},
  \citenamefont {Ouyang}, \citenamefont {Widom},\ and\ \citenamefont
  {Hawk}}]{Gao_COSMS2017}%
  \BibitemOpen
  \bibfield  {author} {\bibinfo {author} {\bibfnamefont {M.}~\bibnamefont
  {Gao}}, \bibinfo {author} {\bibfnamefont {C.}~\bibnamefont {Zhang}}, \bibinfo
  {author} {\bibfnamefont {P.}~\bibnamefont {Gao}}, \bibinfo {author}
  {\bibfnamefont {F.}~\bibnamefont {Zhang}}, \bibinfo {author} {\bibfnamefont
  {L.}~\bibnamefont {Ouyang}}, \bibinfo {author} {\bibfnamefont
  {M.}~\bibnamefont {Widom}}, \ and\ \bibinfo {author} {\bibfnamefont
  {J.}~\bibnamefont {Hawk}},\ }\href {\doibase
  https://doi.org/10.1016/j.cossms.2017.08.001} {\bibfield  {journal} {\bibinfo
   {journal} {Current Opinion in Solid State and Materials Science}\ }\textbf
  {\bibinfo {volume} {21}},\ \bibinfo {pages} {238} (\bibinfo {year}
  {2017})}\BibitemShut {NoStop}%
\bibitem [{\citenamefont {Tan}\ \emph {et~al.}(2018)\citenamefont {Tan},
  \citenamefont {Li}, \citenamefont {Tang}, \citenamefont {Wang},\ and\
  \citenamefont {Kou}}]{Tan_JAC2018430}%
  \BibitemOpen
  \bibfield  {author} {\bibinfo {author} {\bibfnamefont {Y.}~\bibnamefont
  {Tan}}, \bibinfo {author} {\bibfnamefont {J.}~\bibnamefont {Li}}, \bibinfo
  {author} {\bibfnamefont {Z.}~\bibnamefont {Tang}}, \bibinfo {author}
  {\bibfnamefont {J.}~\bibnamefont {Wang}}, \ and\ \bibinfo {author}
  {\bibfnamefont {H.}~\bibnamefont {Kou}},\ }\href {\doibase
  https://doi.org/10.1016/j.jallcom.2018.01.252} {\bibfield  {journal}
  {\bibinfo  {journal} {Journal of Alloys and Compounds}\ }\textbf {\bibinfo
  {volume} {742}},\ \bibinfo {pages} {430} (\bibinfo {year}
  {2018})}\BibitemShut {NoStop}%
\bibitem [{\citenamefont {Ye}\ \emph {et~al.}(2016)\citenamefont {Ye},
  \citenamefont {Wang}, \citenamefont {Lu}, \citenamefont {Liu},\ and\
  \citenamefont {Yang}}]{Ye_MT2016349}%
  \BibitemOpen
  \bibfield  {author} {\bibinfo {author} {\bibfnamefont {Y.}~\bibnamefont
  {Ye}}, \bibinfo {author} {\bibfnamefont {Q.}~\bibnamefont {Wang}}, \bibinfo
  {author} {\bibfnamefont {J.}~\bibnamefont {Lu}}, \bibinfo {author}
  {\bibfnamefont {C.}~\bibnamefont {Liu}}, \ and\ \bibinfo {author}
  {\bibfnamefont {Y.}~\bibnamefont {Yang}},\ }\href {\doibase
  https://doi.org/10.1016/j.mattod.2015.11.026} {\bibfield  {journal} {\bibinfo
   {journal} {Materials Today}\ }\textbf {\bibinfo {volume} {19}},\ \bibinfo
  {pages} {349} (\bibinfo {year} {2016})}\BibitemShut {NoStop}%
\bibitem [{\citenamefont {Borg}\ \emph {et~al.}(2020)\citenamefont {Borg},
  \citenamefont {Frey}, \citenamefont {Moh}, \citenamefont {Pollock},
  \citenamefont {Gorsse}, \citenamefont {Miracle}, \citenamefont {Senkov},
  \citenamefont {Meredig},\ and\ \citenamefont {Saal}}]{borg_SciData_2020}%
  \BibitemOpen
  \bibfield  {author} {\bibinfo {author} {\bibfnamefont {C.~K.~H.}\
  \bibnamefont {Borg}}, \bibinfo {author} {\bibfnamefont {C.}~\bibnamefont
  {Frey}}, \bibinfo {author} {\bibfnamefont {J.}~\bibnamefont {Moh}}, \bibinfo
  {author} {\bibfnamefont {T.~M.}\ \bibnamefont {Pollock}}, \bibinfo {author}
  {\bibfnamefont {S.}~\bibnamefont {Gorsse}}, \bibinfo {author} {\bibfnamefont
  {D.~B.}\ \bibnamefont {Miracle}}, \bibinfo {author} {\bibfnamefont {O.~N.}\
  \bibnamefont {Senkov}}, \bibinfo {author} {\bibfnamefont {B.}~\bibnamefont
  {Meredig}}, \ and\ \bibinfo {author} {\bibfnamefont {J.~E.}\ \bibnamefont
  {Saal}},\ }\href {\doibase 10.1038/s41597-020-00768-9} {\bibfield  {journal}
  {\bibinfo  {journal} {Scientific Data}\ }\textbf {\bibinfo {volume} {7}},\
  \bibinfo {pages} {430} (\bibinfo {year} {2020})}\BibitemShut {NoStop}%
\bibitem [{\citenamefont {Ward}\ \emph {et~al.}(2016)\citenamefont {Ward},
  \citenamefont {Agrawal}, \citenamefont {Choudhary},\ and\ \citenamefont
  {Wolverton}}]{ward_general-purpose_2016}%
  \BibitemOpen
  \bibfield  {author} {\bibinfo {author} {\bibfnamefont {L.}~\bibnamefont
  {Ward}}, \bibinfo {author} {\bibfnamefont {A.}~\bibnamefont {Agrawal}},
  \bibinfo {author} {\bibfnamefont {A.}~\bibnamefont {Choudhary}}, \ and\
  \bibinfo {author} {\bibfnamefont {C.}~\bibnamefont {Wolverton}},\ }\href
  {\doibase 10.1038/npjcompumats.2016.28} {\bibfield  {journal} {\bibinfo
  {journal} {npj Computational Materials}\ }\textbf {\bibinfo {volume} {2}},\
  \bibinfo {pages} {16028} (\bibinfo {year} {2016})}\BibitemShut {NoStop}%
\bibitem [{\citenamefont {Takeuchi}\ and\ \citenamefont
  {Inoue}(2005{\natexlab{a}})}]{miedema_2005}%
  \BibitemOpen
  \bibfield  {author} {\bibinfo {author} {\bibfnamefont {A.}~\bibnamefont
  {Takeuchi}}\ and\ \bibinfo {author} {\bibfnamefont {A.}~\bibnamefont
  {Inoue}},\ }\href {\doibase 10.2320/matertrans.46.2817} {\bibfield  {journal}
  {\bibinfo  {journal} {MATERIALS TRANSACTIONS}\ }\textbf {\bibinfo {volume}
  {46}},\ \bibinfo {pages} {2817} (\bibinfo {year}
  {2005}{\natexlab{a}})}\BibitemShut {NoStop}%
\bibitem [{\citenamefont {John D.~Kelleher}(2020)}]{PCC_2020}%
  \BibitemOpen
  \bibfield  {author} {\bibinfo {author} {\bibfnamefont {A.~D.}\ \bibnamefont
  {John D.~Kelleher}, \bibfnamefont {Brian Mac~Namee}},\ }\href@noop {} {\emph
  {\bibinfo {title} {Fundamentals of machine learning for predictive data
  analytics: algorithms, worked examples, and case studies}}}\ (\bibinfo
  {publisher} {The MIT Press},\ \bibinfo {year} {2020})\BibitemShut {NoStop}%
\bibitem [{\citenamefont {Barraza}\ \emph {et~al.}(2019)\citenamefont
  {Barraza}, \citenamefont {Moro}, \citenamefont {Ferreyra},\ and\
  \citenamefont {de~la Pe\~{n}a}}]{MI1177}%
  \BibitemOpen
  \bibfield  {author} {\bibinfo {author} {\bibfnamefont {N.}~\bibnamefont
  {Barraza}}, \bibinfo {author} {\bibfnamefont {S.}~\bibnamefont {Moro}},
  \bibinfo {author} {\bibfnamefont {M.}~\bibnamefont {Ferreyra}}, \ and\
  \bibinfo {author} {\bibfnamefont {A.}~\bibnamefont {de~la Pe\~{n}a}},\ }\href
  {\doibase 10.1177/0165551518770967} {\bibfield  {journal} {\bibinfo
  {journal} {Journal of Information Science}\ }\textbf {\bibinfo {volume}
  {45}},\ \bibinfo {pages} {53} (\bibinfo {year} {2019})}\BibitemShut {NoStop}%
\bibitem [{\citenamefont {Estevez}\ \emph {et~al.}(2009)\citenamefont
  {Estevez}, \citenamefont {Tesmer}, \citenamefont {Perez},\ and\ \citenamefont
  {Zurada}}]{NMI4749258}%
  \BibitemOpen
  \bibfield  {author} {\bibinfo {author} {\bibfnamefont {P.~A.}\ \bibnamefont
  {Estevez}}, \bibinfo {author} {\bibfnamefont {M.}~\bibnamefont {Tesmer}},
  \bibinfo {author} {\bibfnamefont {C.~A.}\ \bibnamefont {Perez}}, \ and\
  \bibinfo {author} {\bibfnamefont {J.~M.}\ \bibnamefont {Zurada}},\ }\href
  {\doibase 10.1109/TNN.2008.2005601} {\bibfield  {journal} {\bibinfo
  {journal} {IEEE Transactions on Neural Networks}\ }\textbf {\bibinfo {volume}
  {20}},\ \bibinfo {pages} {189} (\bibinfo {year} {2009})}\BibitemShut
  {NoStop}%
\bibitem [{\citenamefont {Biecek}, \citenamefont {Maksymiuk},\ and\
  \citenamefont {Baniecki}(2021)}]{Dalexpackage}%
  \BibitemOpen
  \bibfield  {author} {\bibinfo {author} {\bibfnamefont {P.}~\bibnamefont
  {Biecek}}, \bibinfo {author} {\bibfnamefont {S.}~\bibnamefont {Maksymiuk}}, \
  and\ \bibinfo {author} {\bibfnamefont {H.}~\bibnamefont {Baniecki}},\ }\href
  {https://dalex.drwhy.ai, https://github.com/ModelOriented/DALEX} {\emph
  {\bibinfo {title} {moDel Agnostic Language for Exploration and eXplanation}}}
  (\bibinfo {year} {2021}),\ \bibinfo {note} {{R package version
  2.2.0}}\BibitemShut {NoStop}%
\bibitem [{\citenamefont {Staniak}\ and\ \citenamefont
  {Biecek}(2018)}]{RJ-2018-072}%
  \BibitemOpen
  \bibfield  {author} {\bibinfo {author} {\bibfnamefont {M.}~\bibnamefont
  {Staniak}}\ and\ \bibinfo {author} {\bibfnamefont {P.}~\bibnamefont
  {Biecek}},\ }\href {\doibase 10.32614/RJ-2018-072} {\bibfield  {journal}
  {\bibinfo  {journal} {{The R Journal}}\ }\textbf {\bibinfo {volume} {10}},\
  \bibinfo {pages} {395} (\bibinfo {year} {2018})}\BibitemShut {NoStop}%
\bibitem [{\citenamefont {Takeuchi}\ and\ \citenamefont
  {Inoue}(2000)}]{20001372Takeuchi}%
  \BibitemOpen
  \bibfield  {author} {\bibinfo {author} {\bibfnamefont {A.}~\bibnamefont
  {Takeuchi}}\ and\ \bibinfo {author} {\bibfnamefont {A.}~\bibnamefont
  {Inoue}},\ }\href {\doibase 10.2320/matertrans1989.41.1372} {\bibfield
  {journal} {\bibinfo  {journal} {Materials Transactions, JIM}\ }\textbf
  {\bibinfo {volume} {41}},\ \bibinfo {pages} {1372} (\bibinfo {year}
  {2000})}\BibitemShut {NoStop}%
\bibitem [{\citenamefont {Takeuchi}\ and\ \citenamefont
  {Inoue}(2005{\natexlab{b}})}]{2005Takeuchi}%
  \BibitemOpen
  \bibfield  {author} {\bibinfo {author} {\bibfnamefont {A.}~\bibnamefont
  {Takeuchi}}\ and\ \bibinfo {author} {\bibfnamefont {A.}~\bibnamefont
  {Inoue}},\ }\href {\doibase 10.2320/matertrans.46.2817} {\bibfield  {journal}
  {\bibinfo  {journal} {MATERIALS TRANSACTIONS}\ }\textbf {\bibinfo {volume}
  {46}},\ \bibinfo {pages} {2817} (\bibinfo {year}
  {2005}{\natexlab{b}})}\BibitemShut {NoStop}%
\bibitem [{\citenamefont {Zhang}\ \emph {et~al.}(2008)\citenamefont {Zhang},
  \citenamefont {Zhou}, \citenamefont {Lin}, \citenamefont {Chen},\ and\
  \citenamefont {Liaw}}]{200700240Zhang}%
  \BibitemOpen
  \bibfield  {author} {\bibinfo {author} {\bibfnamefont {Y.}~\bibnamefont
  {Zhang}}, \bibinfo {author} {\bibfnamefont {Y.}~\bibnamefont {Zhou}},
  \bibinfo {author} {\bibfnamefont {J.}~\bibnamefont {Lin}}, \bibinfo {author}
  {\bibfnamefont {G.}~\bibnamefont {Chen}}, \ and\ \bibinfo {author}
  {\bibfnamefont {P.}~\bibnamefont {Liaw}},\ }\href {\doibase
  https://doi.org/10.1002/adem.200700240} {\bibfield  {journal} {\bibinfo
  {journal} {Advanced Engineering Materials}\ }\textbf {\bibinfo {volume}
  {10}},\ \bibinfo {pages} {534} (\bibinfo {year} {2008})},\ \Eprint
  {http://arxiv.org/abs/https://onlinelibrary.wiley.com/doi/pdf/10.1002/adem.200700240}
  {https://onlinelibrary.wiley.com/doi/pdf/10.1002/adem.200700240} \BibitemShut
  {NoStop}%
\bibitem [{\citenamefont {Ling}\ \emph {et~al.}(2017)\citenamefont {Ling},
  \citenamefont {Hutchinson}, \citenamefont {Antono}, \citenamefont {DeCost},
  \citenamefont {Holm},\ and\ \citenamefont {Meredig}}]{ling2017building}%
  \BibitemOpen
  \bibfield  {author} {\bibinfo {author} {\bibfnamefont {J.}~\bibnamefont
  {Ling}}, \bibinfo {author} {\bibfnamefont {M.}~\bibnamefont {Hutchinson}},
  \bibinfo {author} {\bibfnamefont {E.}~\bibnamefont {Antono}}, \bibinfo
  {author} {\bibfnamefont {B.}~\bibnamefont {DeCost}}, \bibinfo {author}
  {\bibfnamefont {E.~A.}\ \bibnamefont {Holm}}, \ and\ \bibinfo {author}
  {\bibfnamefont {B.}~\bibnamefont {Meredig}},\ }\href@noop {} {\bibfield
  {journal} {\bibinfo  {journal} {Materials Discovery}\ }\textbf {\bibinfo
  {volume} {10}},\ \bibinfo {pages} {19} (\bibinfo {year} {2017})}\BibitemShut
  {NoStop}%
\bibitem [{\citenamefont {Xie}\ and\ \citenamefont
  {Grossman}(2018)}]{xie2018crystal}%
  \BibitemOpen
  \bibfield  {author} {\bibinfo {author} {\bibfnamefont {T.}~\bibnamefont
  {Xie}}\ and\ \bibinfo {author} {\bibfnamefont {J.~C.}\ \bibnamefont
  {Grossman}},\ }\href@noop {} {\bibfield  {journal} {\bibinfo  {journal}
  {Physical review letters}\ }\textbf {\bibinfo {volume} {120}},\ \bibinfo
  {pages} {145301} (\bibinfo {year} {2018})}\BibitemShut {NoStop}%
\bibitem [{\citenamefont {Lopez}\ \emph {et~al.}(2017)\citenamefont {Lopez},
  \citenamefont {Sanchez-Lengeling}, \citenamefont {de~Goes~Soares},\ and\
  \citenamefont {Aspuru-Guzik}}]{lopez2017design}%
  \BibitemOpen
  \bibfield  {author} {\bibinfo {author} {\bibfnamefont {S.~A.}\ \bibnamefont
  {Lopez}}, \bibinfo {author} {\bibfnamefont {B.}~\bibnamefont
  {Sanchez-Lengeling}}, \bibinfo {author} {\bibfnamefont {J.}~\bibnamefont
  {de~Goes~Soares}}, \ and\ \bibinfo {author} {\bibfnamefont {A.}~\bibnamefont
  {Aspuru-Guzik}},\ }\href@noop {} {\bibfield  {journal} {\bibinfo  {journal}
  {Joule}\ }\textbf {\bibinfo {volume} {1}},\ \bibinfo {pages} {857} (\bibinfo
  {year} {2017})}\BibitemShut {NoStop}%
\bibitem [{\citenamefont {Gurnani}\ \emph {et~al.}(2021)\citenamefont
  {Gurnani}, \citenamefont {Yu}, \citenamefont {Kim}, \citenamefont {Sholl},\
  and\ \citenamefont {Ramprasad}}]{Rampi_SHAP}%
  \BibitemOpen
  \bibfield  {author} {\bibinfo {author} {\bibfnamefont {R.}~\bibnamefont
  {Gurnani}}, \bibinfo {author} {\bibfnamefont {Z.}~\bibnamefont {Yu}},
  \bibinfo {author} {\bibfnamefont {C.}~\bibnamefont {Kim}}, \bibinfo {author}
  {\bibfnamefont {D.~S.}\ \bibnamefont {Sholl}}, \ and\ \bibinfo {author}
  {\bibfnamefont {R.}~\bibnamefont {Ramprasad}},\ }\href {\doibase
  10.1021/acs.chemmater.0c04729} {\bibfield  {journal} {\bibinfo  {journal}
  {Chemistry of Materials}\ }\textbf {\bibinfo {volume} {33}},\ \bibinfo
  {pages} {3543} (\bibinfo {year} {2021})}\BibitemShut {NoStop}%
\bibitem [{\citenamefont {Xiong}, \citenamefont {Shi},\ and\ \citenamefont
  {Zhang}(2021)}]{XIONG2021133}%
  \BibitemOpen
  \bibfield  {author} {\bibinfo {author} {\bibfnamefont {J.}~\bibnamefont
  {Xiong}}, \bibinfo {author} {\bibfnamefont {S.-Q.}\ \bibnamefont {Shi}}, \
  and\ \bibinfo {author} {\bibfnamefont {T.-Y.}\ \bibnamefont {Zhang}},\ }\href
  {\doibase https://doi.org/10.1016/j.jmst.2021.01.054} {\bibfield  {journal}
  {\bibinfo  {journal} {Journal of Materials Science \& Technology}\ }\textbf
  {\bibinfo {volume} {87}},\ \bibinfo {pages} {133} (\bibinfo {year}
  {2021})}\BibitemShut {NoStop}%
\bibitem [{\citenamefont {Clough}\ \emph {et~al.}(2019)\citenamefont {Clough},
  \citenamefont {Oksuz}, \citenamefont {Puyol-Ant{\'o}n}, \citenamefont
  {Ruijsink}, \citenamefont {King},\ and\ \citenamefont
  {Schnabel}}]{clough2019global}%
  \BibitemOpen
  \bibfield  {author} {\bibinfo {author} {\bibfnamefont {J.~R.}\ \bibnamefont
  {Clough}}, \bibinfo {author} {\bibfnamefont {I.}~\bibnamefont {Oksuz}},
  \bibinfo {author} {\bibfnamefont {E.}~\bibnamefont {Puyol-Ant{\'o}n}},
  \bibinfo {author} {\bibfnamefont {B.}~\bibnamefont {Ruijsink}}, \bibinfo
  {author} {\bibfnamefont {A.~P.}\ \bibnamefont {King}}, \ and\ \bibinfo
  {author} {\bibfnamefont {J.~A.}\ \bibnamefont {Schnabel}},\ }in\ \href@noop
  {} {\emph {\bibinfo {booktitle} {International Conference on Medical Image
  Computing and Computer-Assisted Intervention}}}\ (\bibinfo {organization}
  {Springer},\ \bibinfo {year} {2019})\ pp.\ \bibinfo {pages}
  {656--664}\BibitemShut {NoStop}%
\bibitem [{\citenamefont {George}, \citenamefont {Curtin},\ and\ \citenamefont
  {Tasan}(2020)}]{GEORGE2020435}%
  \BibitemOpen
  \bibfield  {author} {\bibinfo {author} {\bibfnamefont {E.}~\bibnamefont
  {George}}, \bibinfo {author} {\bibfnamefont {W.}~\bibnamefont {Curtin}}, \
  and\ \bibinfo {author} {\bibfnamefont {C.}~\bibnamefont {Tasan}},\ }\href
  {\doibase https://doi.org/10.1016/j.actamat.2019.12.015} {\bibfield
  {journal} {\bibinfo  {journal} {Acta Materialia}\ }\textbf {\bibinfo {volume}
  {188}},\ \bibinfo {pages} {435} (\bibinfo {year} {2020})}\BibitemShut
  {NoStop}%
\bibitem [{\citenamefont {Chiquet}\ \emph {et~al.}(2020)\citenamefont
  {Chiquet}, \citenamefont {Rigaill}, \citenamefont {Sundqvist},\ and\
  \citenamefont {Dervieux}}]{aricode}%
  \BibitemOpen
  \bibfield  {author} {\bibinfo {author} {\bibfnamefont {J.}~\bibnamefont
  {Chiquet}}, \bibinfo {author} {\bibfnamefont {G.}~\bibnamefont {Rigaill}},
  \bibinfo {author} {\bibfnamefont {M.}~\bibnamefont {Sundqvist}}, \ and\
  \bibinfo {author} {\bibfnamefont {V.}~\bibnamefont {Dervieux}},\ }\href
  {https://github.com/jchiquet/aricode} {\emph {\bibinfo {title} {{aricode:
  Efficient Computations of Standard Clustering Comparison Measures}}}}
  (\bibinfo {year} {2020}),\ \bibinfo {note} {{R package version
  1.0.0}}\BibitemShut {NoStop}%
\bibitem [{\citenamefont {{R Core Team}}(2012)}]{RSTUDIO}%
  \BibitemOpen
  \bibfield  {author} {\bibinfo {author} {\bibnamefont {{R Core Team}}},\
  }\href {http://www.R-project.org/} {\emph {\bibinfo {title} {{R: A Language
  and Environment for Statistical Computing}}}},\ \bibinfo {organization} {R
  Foundation for Statistical Computing},\ \bibinfo {address} {Vienna, Austria}
  (\bibinfo {year} {2012}),\ \bibinfo {note} {{ISBN} 3-900051-07-0}\BibitemShut
  {NoStop}%
\bibitem [{\citenamefont {Vapnik}(2000)}]{SVM}%
  \BibitemOpen
  \bibfield  {author} {\bibinfo {author} {\bibfnamefont {V.}~\bibnamefont
  {Vapnik}},\ }\href {\doibase 10.1007/978-1-4757-3264-1} {\emph {\bibinfo
  {title} {{The Nature of Statistical Learning Theory}}}}\ (\bibinfo
  {publisher} {Springer-Verlag New York},\ \bibinfo {year} {2000})\BibitemShut
  {NoStop}%
\bibitem [{\citenamefont {MacKinnon}, \citenamefont {Lockwood},\ and\
  \citenamefont {Williams}(2004)}]{Bootstrap_resampling}%
  \BibitemOpen
  \bibfield  {author} {\bibinfo {author} {\bibfnamefont {D.~P.}\ \bibnamefont
  {MacKinnon}}, \bibinfo {author} {\bibfnamefont {C.~M.}\ \bibnamefont
  {Lockwood}}, \ and\ \bibinfo {author} {\bibfnamefont {J.}~\bibnamefont
  {Williams}},\ }\href {\doibase 10.1207/s15327906mbr3901\_4} {\bibfield
  {journal} {\bibinfo  {journal} {Multivariate Behavioral Research}\ }\textbf
  {\bibinfo {volume} {39}},\ \bibinfo {pages} {99} (\bibinfo {year}
  {2004})}\BibitemShut {NoStop}%
\bibitem [{\citenamefont {Meyer}\ \emph {et~al.}(2015)\citenamefont {Meyer},
  \citenamefont {Dimitriadou}, \citenamefont {Hornik}, \citenamefont
  {Weingessel},\ and\ \citenamefont {Leisch}}]{e1071}%
  \BibitemOpen
  \bibfield  {author} {\bibinfo {author} {\bibfnamefont {D.}~\bibnamefont
  {Meyer}}, \bibinfo {author} {\bibfnamefont {E.}~\bibnamefont {Dimitriadou}},
  \bibinfo {author} {\bibfnamefont {K.}~\bibnamefont {Hornik}}, \bibinfo
  {author} {\bibfnamefont {A.}~\bibnamefont {Weingessel}}, \ and\ \bibinfo
  {author} {\bibfnamefont {F.}~\bibnamefont {Leisch}},\ }\href
  {http://CRAN.R-project.org/package=e1071} {\emph {\bibinfo {title} {{e1071:
  Misc Functions of the Department of Statistics, Probability Theory Group
  (Formerly: E1071), TU Wien}}}} (\bibinfo {year} {2015}),\ \bibinfo {note} {{R
  package version 1.6-7}}\BibitemShut {NoStop}%
\bibitem [{\citenamefont {Kassambara}\ and\ \citenamefont
  {Mundt}(2020)}]{kmeans_package}%
  \BibitemOpen
  \bibfield  {author} {\bibinfo {author} {\bibfnamefont {A.}~\bibnamefont
  {Kassambara}}\ and\ \bibinfo {author} {\bibfnamefont {F.}~\bibnamefont
  {Mundt}},\ }\href {http://www.sthda.com/english/rpkgs/factoextra} {\emph
  {\bibinfo {title} {Extract and Visualize the Results of Multivariate Data
  Analyses}}} (\bibinfo {year} {2020}),\ \bibinfo {note} {{R package version
  1.0.7}}\BibitemShut {NoStop}%
\bibitem [{\citenamefont {Chang}\ \emph {et~al.}(2020)\citenamefont {Chang},
  \citenamefont {Cheng}, \citenamefont {Allaire}, \citenamefont {Xie},\ and\
  \citenamefont {McPherson}}]{RShiny}%
  \BibitemOpen
  \bibfield  {author} {\bibinfo {author} {\bibfnamefont {W.}~\bibnamefont
  {Chang}}, \bibinfo {author} {\bibfnamefont {J.}~\bibnamefont {Cheng}},
  \bibinfo {author} {\bibfnamefont {J.}~\bibnamefont {Allaire}}, \bibinfo
  {author} {\bibfnamefont {Y.}~\bibnamefont {Xie}}, \ and\ \bibinfo {author}
  {\bibfnamefont {J.}~\bibnamefont {McPherson}},\ }\href
  {https://CRAN.R-project.org/package=shiny} {\emph {\bibinfo {title} {{shiny:
  Web Application Framework for R}}}} (\bibinfo {year} {2020}),\ \bibinfo
  {note} {r package version 1.5.0}\BibitemShut {NoStop}%
\bibitem [{\citenamefont {Lee}\ \emph {et~al.}(2021{\natexlab{b}})\citenamefont
  {Lee}, \citenamefont {Ayyasamy}, \citenamefont {Delsa}, \citenamefont
  {Hartnett},\ and\ \citenamefont {Balachandran}}]{FigshareDataset}%
  \BibitemOpen
  \bibfield  {author} {\bibinfo {author} {\bibfnamefont {K.}~\bibnamefont
  {Lee}}, \bibinfo {author} {\bibfnamefont {M.~V.}\ \bibnamefont {Ayyasamy}},
  \bibinfo {author} {\bibfnamefont {P.}~\bibnamefont {Delsa}}, \bibinfo
  {author} {\bibfnamefont {T.~Q.}\ \bibnamefont {Hartnett}}, \ and\ \bibinfo
  {author} {\bibfnamefont {P.~V.}\ \bibnamefont {Balachandran}},\ }\href
  {\doibase https://doi.org/10.6084/m9.figshare.15098094.v1} {\bibfield
  {journal} {\bibinfo  {journal} {figshare}\ } (\bibinfo {year}
  {2021}{\natexlab{b}}),\
  https://doi.org/10.6084/m9.figshare.15098094.v1}\BibitemShut {NoStop}%
\end{thebibliography}%
%

 \end{document}